\documentclass[aps,prb,notitlepage,nofootinbib,superscriptaddress]{revtex4-2}
\usepackage[utf8]{inputenc}
\usepackage{amsmath,amssymb,amsfonts,stmaryrd}
\usepackage{physics}
\usepackage{dsfont}
\usepackage{graphicx}
\usepackage{xcolor}
\usepackage{graphicx}
\usepackage{mathtools}
\usepackage{comment}
\usepackage{cancel}
\usepackage{hyperref}
\hypersetup{
    colorlinks=true, 
    linktoc=all,     
    linkcolor=blue,  
}

\usepackage{bm} 
\usepackage{subfigure} 
\usepackage{environ}
\usepackage{url}
\usepackage{hyperref}
\usepackage[margin=0.75in]{geometry}
\usepackage{leftidx}
\usepackage{booktabs}
\usepackage{qcircuit}
\usepackage[scr=boondox]{mathalpha}

\newcommand{\Z}{{\mathbb Z}}

\newcommand{\gsf}{{G_{\mathrm{sf}}}}

\NewEnviron{eqs}{%
\begin{equation}\begin{split}
    \BODY
\end{split}\end{equation}
}

\begin{document}

\title{Soft symmetries of topological orders}
\author{Ryohei Kobayashi}
\affiliation{School of Natural Sciences, Institute for Advanced Study, Princeton, NJ 08540, USA} 
\author{Maissam Barkeshli}
\affiliation{Department of Physics and Joint Quantum Institute, University of Maryland,
College Park, Maryland 20742, USA}

\maketitle

(2+1)D topological orders possess emergent symmetries given by a group $\text{Aut}_{br}(\mathcal{C})$, which consists of the braided tensor autoequivalences of the modular tensor category $\mathcal{C}$ that describes the anyons. In this paper we discuss cases where $\text{Aut}_{br}(\mathcal{C})$ has elements that neither permute anyons nor are associated to any symmetry fractionalization but are still non-trivial, which we refer to as soft symmetries.
We point out that one can construct topological defects corresponding to such exotic symmetry actions by decorating with a certain class of gauged SPT states that cannot be distinguished by their torus partition function. This gives a physical interpretation to work by Davydov on soft braided tensor autoequivalences. This has a number of important implications for the classification of gapped boundaries, non-invertible spontaneous symmetry breaking, and the general classification of symmetry-enriched topological phases of matter. 
We also demonstrate analogous phenomena in higher dimensions, such as (3+1)D gauge theory with gauge group given by the quaternion group $Q_8$.

\bigskip \bigskip \bigskip


\bigskip


\tableofcontents

\unitlength = .8mm

\setcounter{tocdepth}{3}

\bigskip

\section{Introduction}

A fundamental question in quantum many-body physics is to understand the relationship between global symmetries and topological order, which is the subject of \it symmetry-enriched \rm topological phases of matter \cite{wen04,zeng2019quantum,Barkeshli2019}. The most well-known example of a such a relationship is how anyons of a topologically ordered state can carry fractional quantum numbers under a global symmetry, a phenomenon known as symmetry fractionalization, with prominent examples being the fractional electric charges in the fractional quantum Hall effect \cite{laughlin1983,de1997direct,saminadayar1997observation}, or fractional spin in quantum spin liquids \cite{Savary2017}. A more recently understood phenomenon relates to how discrete global symmetries can permute topologically distinct anyon types, which leads to the striking possibility of non-Abelian twist defects, even when the underlying topological phase only has Abelian anyons \cite{barkeshli2010,bombin2010,barkeshli2012,barkeshli2013}. These play an important role in fault-tolerant topological quantum computation \cite{lavasani2018low,brown2017poking,hastings2014reduced,zhu2020quantum}; they have been experimentally demonstrated in quantum computers \cite{google2023non,iqbal2024qutrit} and a variety of experimental proposals have been suggested for realizing them in two-dimensional electron systems \cite{barkeshli2014,barkeshli2014deg,clarke2013,lindner2012,cheng2012,agarwal2023quantum}. 

A longstanding question in this subject has been whether there can be symmetries that are neither fractionalized nor permute anyons, but nevertheless have a non-trivial action on the topological state. We refer to such symmetries as {\bf soft symmetries}. In this paper, we demonstrate how soft symmetries can indeed arise. The key insight that we exploit is that in a (2+1)D topological phase described by $G$ gauge theory, we can decorate a codimension-$1$ submanifold with a (1+1)D $G$-symmetry protected topological state ($G$-SPT) before gauging $G$ everywhere \cite{Barkeshli2023codim2}. The result is an invertible codimension-1 topological defect that we refer to as a gauged SPT defect. Since the gauged SPT defect is invertible, it can be pumped across the entire system by a finite depth local unitary circuit and keeps the ground state subspace invariant. As such, we consider the corresponding circuit to be an \it emergent symmetry \rm of the topological state.\footnote{In our construction, we get a non-onsite emergent symmetry in general.} 
As we will discuss, if the SPT cannot be detected by its torus partition function, then the resulting gauged SPT defect has the surprising property mentioned above, which we verify explicitly using exactly solvable two-dimensional lattice models of quantum doubles. The emergent symmetry that we find acts as the identity on the ground state subspace on a torus while acting non-trivially on the ground state subspace on higher genus surfaces. 

Our result has a number of important implications for our fundamental understanding of topological order:
\begin{itemize}
    \item With respect to the classification of gapped boundaries \cite{Beigi2011boundary,kitaev2012,kapustin2011,Levin2013edge,barkeshli2013defect,cong2017hamiltonian}, our result demonstrates how there can be inequivalent gapped boundaries that have the same set of condensed anyons.  
    \item The above in turn implies that there are \it two distinct \rm (1+1)D gapped phases where a non-invertible Rep($G$) symmetry is spontaneously broken completely. 
    \item With respect to our understanding of SETs, our results demonstrate how there can be subtle types of symmetries that can only be distinguished by their action on the ground state subspace on surfaces of genus $g \geq 2$.
\end{itemize}

The results presented here also provide a physical interpretation of the soft braided tensor auto-equivalences discovered by Davydov in \cite{Davydov_2014}. Our physical construction in terms of gauged SPT defects gives additional insights that we use to discover soft symmetries in (3+1)D topological orders.

This paper is organized as follows. In Sec.~\ref{sec:symmetry} we first review the algebraic theory of anyons and its autoequivalences, and introduce the algebraic description of soft symmetries. In Sec.~\ref{sec:soft} we introduce an example of a soft symmetry defect realized in finite gauge theory in (2+1)D. In Sec.~\ref{sec:double} we introduce a lattice model of the quantum double that has the soft symmetry. In Sec.~\ref{sec:boundary} we discuss an implication of the soft symmetry for gapped boundaries of finite gauge theory, and discuss how it leads to the distinct SSB phases of $\mathrm{Rep}(G)$ symmetry. In Sec.~\ref{sec:3+1d} we discuss the analogue of soft symmetry in higher dimensions, focusing on $Q_8$ gauge theory in (3+1)D. 

\section{Global Symmetries of (2+1)D Topological Order}
\label{sec:symmetry}
Our starting point is a unitary modular tensor category $\mathcal{C}$, which is an algebraic framework to describe the universal fusion and braiding properties of anyons in (2+1)D topological orders. To set the notation, we briefly review the key data below. We refer the reader to Appendix E of \cite{Kitaev2006anyons} and Refs. \cite{bonderson2012non,Barkeshli2019,simon2023topological} for a comprehensive account aimed at physicists. Methods to define this algebraic data in microscopic models were described in \cite{kawagoe2020}.

\subsection{Review: Unitary modular tensor categories }

In the notation below, we follow the convention of \cite{bonderson2008,Barkeshli2019}. A UMTC \cite{moore1989b,wang2008} $\mathcal{C}$ consists of a finite set of anyons $\{a\}$, obeying fusion rules 
\begin{align}
a \times b = \sum_{c} N_{ab}^c c, 
\end{align}
where $N_{ab}^c$ are non-negative integers. The theory contains linear vector spaces, referred to as ``fusion spaces," $V_{ab}^c$, and their dual ``splitting spaces" $V_{c}^{ab}$, such that $\text{dim} V_{ab}^c = \text{dim} V_{c}^{ab} = N_{ab}^c$. The basis states in these spaces are diagrammatically depicted as:

\begin{equation}
\left( d_{c} / d_{a}d_{b} \right) ^{1/4}
\hbox{ \raisebox{-3ex}{\includegraphics[width=1.5cm]{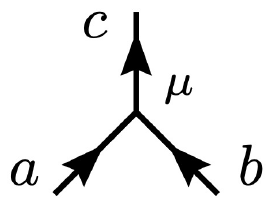}}}
=\left\langle a,b;c,\mu \right| \in
V_{ab}^{c} ,
\hspace{50pt}
\left( d_{c} / d_{a}d_{b}\right) ^{1/4}
\hbox{ \raisebox{-3ex}{\includegraphics[width=1.5cm]{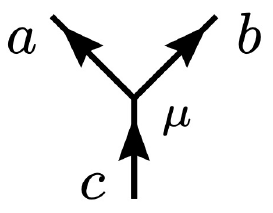}}}
=\left| a,b;c,\mu \right\rangle \in
V_{c}^{ab},
\label{eq:ket}
\end{equation}

where $\mu=1,\ldots ,N_{ab}^{c}$, $d_a$ is the quantum dimension of $a$, 
and the factors $\left(\frac{d_c}{d_a d_b}\right)^{1/4}$ are a normalization convention for the diagrams. The set of anyons contains the identity or vacuum charge $0$, and each anyon $a$ has a dual anti-particle $\bar{a}$ such that $a \times \bar{a} = 0 + \cdots$. 

The theory is defined by a consistent set of $F$ and $R$ symbols, which are unitary maps on the vector spaces:
\begin{align}
    F^{abc}_d : & \bigotimes_e V^{ab}_e \otimes V^{ec}_d \rightarrow \bigotimes_f V^{af}_d \otimes V^{bc}_f ,
    \nonumber \\
    R^{ab}_c : & V^{ba}_c \rightarrow V^{ab}_c . 
\end{align}
These are depicted graphically as 
\begin{equation}
\hbox{ \raisebox{-7ex}{\includegraphics[width=2cm]{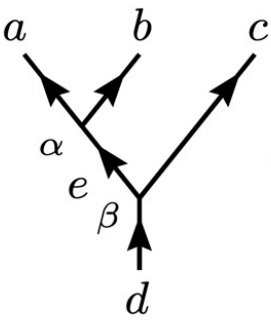}}}
= \sum_{f,\mu,\nu} \left[F_d^{abc}\right]_{(e,\alpha,\beta)(f,\mu,\nu)}
\hbox{ \raisebox{-7ex}{\includegraphics[width=2.1cm]{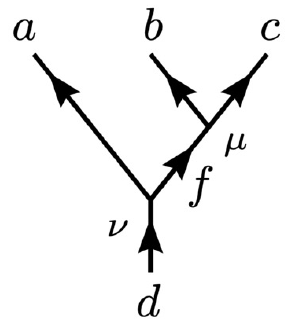}}}
,
\hspace{60pt}
\hbox{ \raisebox{-3.5ex}{\includegraphics[width=1.5cm]{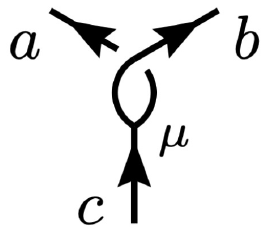}}}
=\sum\limits_{\nu }\left[ R_{c}^{ab}\right] _{\mu \nu}
\hbox{ \raisebox{-3.5ex}{\includegraphics[width=1.6cm]{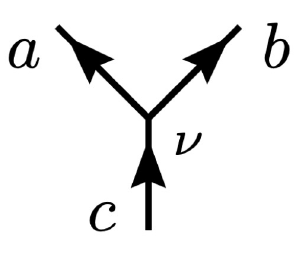}}}
\end{equation}

If we perform a basis transformation, referred to as a vertex basis gauge transformation, on the fusion and splitting states, $F$ and $R$ symbols transform:
\begin{align}
    \Gamma^{ab}_c : &V^{ab}_c \rightarrow V^{ab}_c
    \nonumber \\
    &|a,b;c,\mu\rangle \rightarrow \sum_\nu (\Gamma^{ab}_c)_{\mu \nu} |a,b;c,\nu \rangle
    \nonumber \\
    & (F^{abc}_d)_{ef} \rightarrow \Gamma^{ab}_e \Gamma^{ec}_d (F^{abc}_d)_{ef} (\Gamma^{af}_d)^{-1} (\Gamma^{bc}_f)^{-1}
    \nonumber \\
    & R^{ab}_c \rightarrow \Gamma^{ba}_c R^{ab}_c (\Gamma^{ab}_c)^{-1},
    \label{eq:vertexgauge}
\end{align}
where we have suppressed additional indices in the transformation of the $F$ and $R$ symbols. The terminology ``gauge transformation'' originates from the fact that the matrices $\{\Gamma^{ab}_c\}$ redefine the $F,R$ symbols of the modular tensor category, giving a different expression of the same theory, thus putting a redundancy in the presentation of the theory.

Properties of the UMTC that are invariant under the vertex basis gauge transformation include the topological twists $\theta_a$ and the modular $S$ matrix. A special role will be played by \it natural isomorphisms\rm, which are vertex basis gauge transformations such that 
\begin{align}
(\Gamma^{ab}_c)_{\mu\nu} = \frac{\gamma_a \gamma_b}{\gamma_c} \delta_{\mu \nu},
\end{align}
where $\gamma_a$ is a $U(1)$ phase. These natural isomorphisms do not change the $F$ and $R$ symbols, and leaves all closed anyon diagrams with junctions invariant. Therefore the natural isomorphisms are regarded as trivial gauge transformations, putting a redundancy in the expressions of vertex basis gauge transformations. 

\subsection{0-form symmetry: Braided tensor auto-equivalences}

We define an invertible map $\varphi: \mathcal{C} \rightarrow \mathcal{C}$, which has an action on all of the data of the theory. $\varphi$ is a \it braided tensor autoequivalence \rm if the following is satisfied. It transforms the anyons as
\begin{align}
    a \rightarrow \varphi(a) = a', 
\end{align}
such that gauge invariant quantities are invariant under $\varphi$:
\begin{align}
 N_{a'b'}^{c'} &= N_{ab}^c
\nonumber \\
S_{a'b'} &= S_{ab}
\nonumber \\
\theta_{a'} &= \theta_a .
\end{align}
$\varphi$ acts on the fusion and splitting spaces as a linear map:\footnote{It is possible to generalize to anti-linear operations as well, as in \cite{Barkeshli2019}.}
\begin{align}
    \varphi: V_{c}^{ab} &\rightarrow V_{c'}^{a'b'},
    \nonumber \\
    \varphi(|a,b;c,\mu\rangle) &= \sum_\nu [U_\varphi(a',b';c')]_{\mu \nu} |a',b';c'\nu\rangle
\end{align}
and it transforms the $F$ and $R$ symbols of the theory as follows:
\begin{align}
    \varphi(R^{ab}_c) &= \tilde{R}^{a'b'}_{c'} \equiv U_\varphi(b',a';c') R^{a'b'}_{c'} U_\varphi^\dagger(a',b';c')
    \nonumber \\
    \varphi([F^{abc}_d]_{ef}) &= [\tilde{F}^{a'b'c'}_{d'}]_{e'f'} \equiv U_\varphi(a',b';e') U_\varphi(e',c';f') [F^{a'b'c'}_{d'}]_{e'f'} U_\varphi^\dagger(a',f';d') U_\varphi^\dagger(b',c';f') , 
\end{align}
where again we have suppressed the additional indices. These transformations must by definition keep the $F$ and $R$ symbols invariant:
\begin{align}
\label{Finvariance}
[\tilde{F}^{a'b'c'}_{d'}]_{e'f'} &= [F^{abc}_d]_{ef} ,
    \\
    \label{Rinvariance}
    \tilde{R}^{a'b'}_{c'} &= R^{ab}_c .
\end{align}
We can depict these transformations graphically by introducing a codimension-1 defect labeled by $\varphi$.
\begin{center}
\vspace{-5pt}
\raisebox{-.2\height}{\includegraphics[width=.65\linewidth]{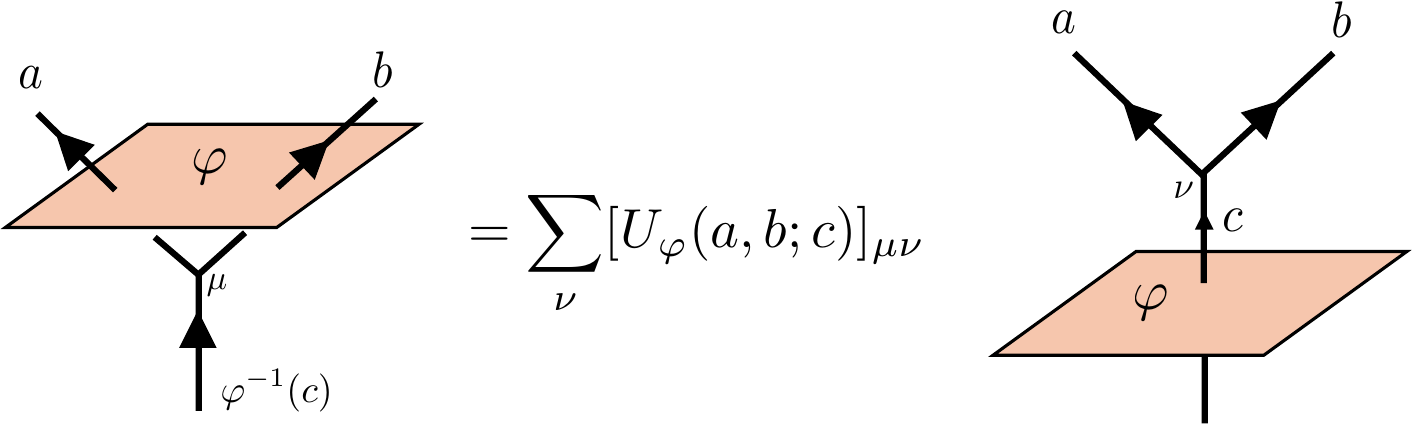}}
\end{center}
The invariance of $F$, Eq. \ref{Finvariance}, can be understood as a consequence of commutativity of the following diagram:
\begin{center}
\vspace{-5pt}
\raisebox{-.2\height}{\includegraphics[width=.65\linewidth]{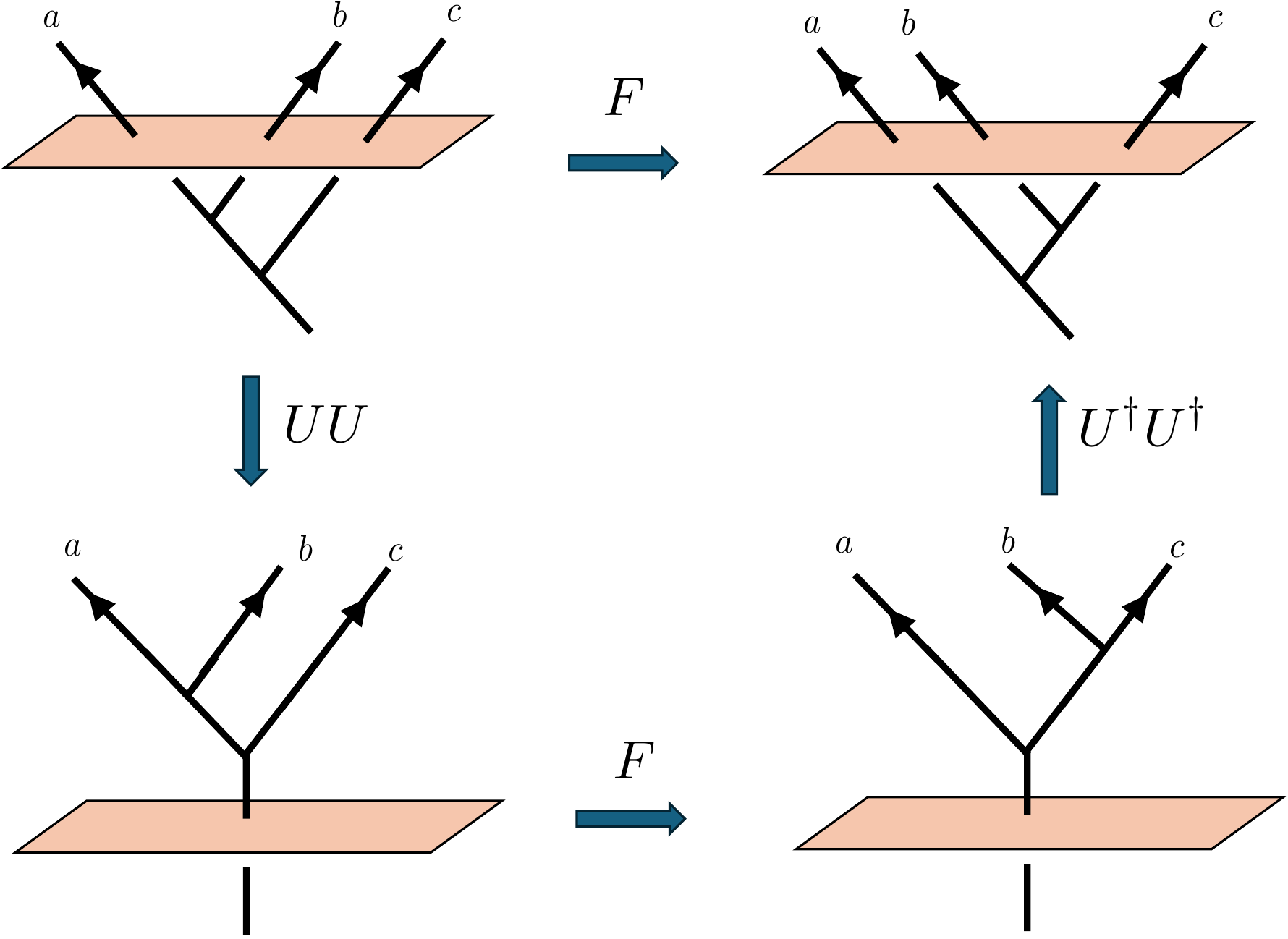}}
\end{center}
The invariance of $R$, Eq. \ref{Rinvariance}, can be derived by introducing additional data associated with $G$-crossed braiding of codefects, as described in Eq. 280 - 282 of \cite{Barkeshli2019}. 

It is natural to consider an equivalence relation on such maps, where two maps $\varphi$ and $\varphi'$ are equivalent if there is a continuous family of maps $\varphi_s$ for $s \in [0,1]$, such that $\varphi_0 = \varphi$ and $\varphi_1 = \varphi'$. Note that $\varphi_s^{-1} \circ \varphi_{s'}$ is a vertex basis gauge transformation that keeps the $F$-symbols invariant 
\begin{align}
   \Gamma^{ab}_e \Gamma^{ec}_f [F^{abc}_{d}]_{ef} (\Gamma^{af}_d)^\dagger (\Gamma^{bc}_f)^\dagger = [F^{abc}_d]_{ef} .
    \label{eq:GammapreservesF}
\end{align}
While such vertex gauge transformations are labeled by continuous unitary matrices on each fusion vertex, up to deformation they generate a finite group. In fact, considering maps $\varphi$ up to continuous deformation is equivalent to considering them up to natural isomorphisms, due to Ocneanu rigidity~\cite{etingof2017fusioncategories,Kitaev2006anyons}. The rigidity states that the infinitesimal gauge transformation $\Gamma^{ab}_c$ satisfying \eqref{eq:GammapreservesF} can be expressed as a natural isomorphism
\begin{align}
    \Gamma^{ab}_c = \frac{\gamma_a \gamma_b}{\gamma_c} \qquad \text{when $\Gamma^{ab}_c\approx \mathrm{id}_{V^{ab}_c}$ and satisfies \eqref{eq:GammapreservesF}}~,
\end{align}
where $\gamma_a,\gamma_b,\gamma_c$ are phase factors close to unity. Taking equivalence classes of braided tensor autoequivalences modulo natural isomorphisms defines a finite group, denoted Aut$_{br}(\mathcal{C})$, which is the group of braided auto-equivalences of $\mathcal{C}$. 

In this paper, we are interested in elements of Aut$_{br}(\mathcal{C})$ that do not permute anyons but which are nevertheless non-trivial, namely $\varphi\in \text{Aut}_{br}(\mathcal{C})$ satisfying the following two conditions:
\begin{itemize}
    \item  $\varphi(a)=a$ for all anyons $a$.
    \item The $U$ symbol  $\{U_\varphi(a,b;c)\}$ is not a natural isomorphism.
\end{itemize}

We refer to such transformations as {\bf soft symmetries} and we denote the group of soft symmetries as Aut$_\text{sf}(\mathcal{C})\subset \text{Aut}_{br}(\mathcal{C})$. In the mathematics literature they are referred to as soft braided tensor autoequivalences. We see that such transformations correspond to vertex basis gauge transformations that keep the $F$ and $R$ symbols invariant and which are not continuously deformable to the identity.

\subsection{Symmetry action on genus $g$ surfaces}

\begin{figure}[t]
    \centering
    \includegraphics[width=0.5\textwidth]{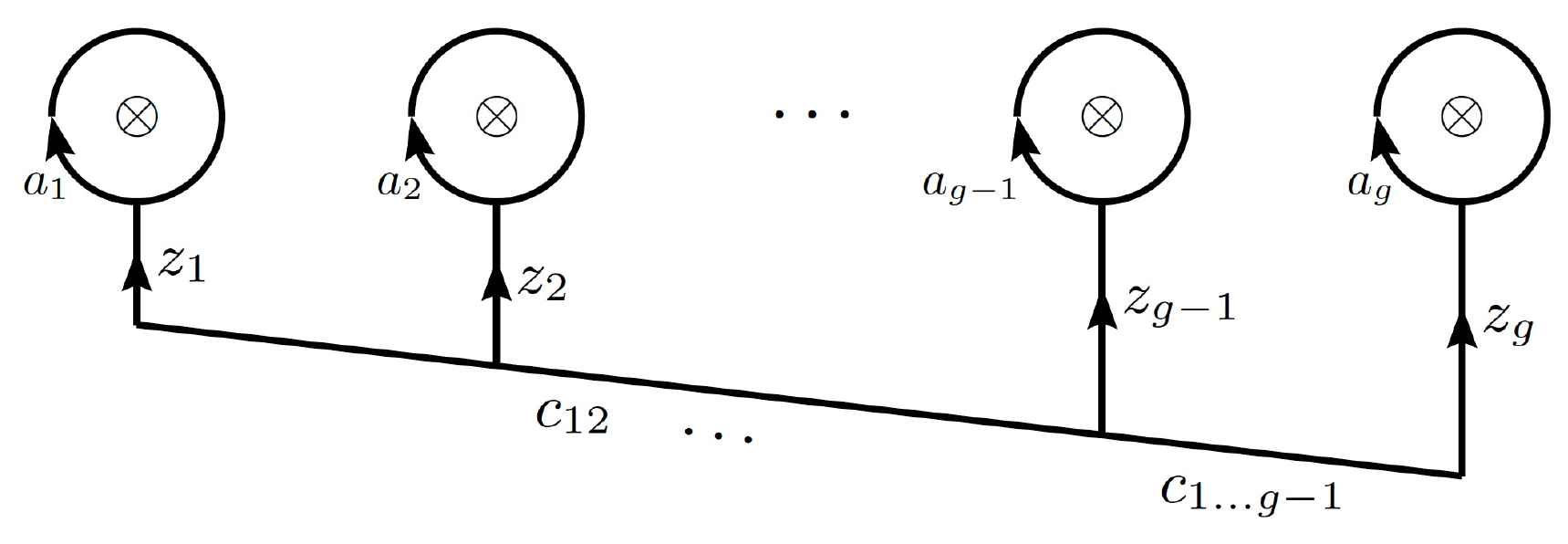}
    \caption{Labels and fusion decomposition defining an orthonormal basis for states on genus $g$ surfaces. The $\otimes$ signifies the fact that the loops are non-contractible on the genus $g$ surface. }
    \label{fig:genusg}
\end{figure}

The $U$ symbols characterize the action of $[\varphi] \in \text{Aut}_{br}(\mathcal{C})$ on the space of states on a closed genus $g$ surface. We can pick an orthonormal basis of states on the genus $g$ surface as
\begin{align}
    \bigotimes_{j=1}^g |a_j,\overline{a_j}; z_j,\mu_j\rangle |c_{1\cdots j-1}, z_j;c_{1\cdots j},\nu_{1\cdots j} \rangle,
\end{align}
with $c_{1\cdots g} = 0$ and $c_1 = z_1$. This corresponds to the diagram shown in Fig.~\ref{fig:genusg}. 
The symmetry action on a state $|\psi\rangle$ is then
\begin{align}
|\psi\rangle \rightarrow \varphi( |\psi \rangle),
\end{align}
with $\varphi$ acting on the basis states to give:
\begin{align}
    \bigotimes_{j=1}^g U_\varphi^{-1}(a_j', \overline{a_j'};0) 
    [U_\varphi(a_j',\overline{a_j'};z'_j)]_{\mu_j', \mu_j} [U_\varphi(c'_{1\cdots j-1}, z'_j;c'_{1\cdots j})]_{\nu_{1\cdots j}', \nu_{1\cdots j}} |a'_j,\overline{a_j'}; z'_j,\mu_j\rangle |c'_{1\cdots j-1}, z'_j;c'_{1\cdots j},\nu_{1\cdots j} \rangle,
\end{align}
with a sum over $\mu_j$ and $\nu_{1\cdots j}$ implied. 

For soft symmetries, the states on a genus $g = 1$ surface transform trivially:
\begin{align}
    \varphi(|a, \overline{a};0 \rangle ) = |a, \bar{a};0\rangle.
\end{align}
For the genus $g = 2$ surface, a soft symmetry transforms the states as
\begin{align}
\varphi(|a_1, \overline{a_1};z_1,\mu_1' \rangle |a_2,\overline{a}_2;\overline{z}_1, \mu_2'\rangle) = 
\frac{ [U_\varphi(a_1,\overline{a}_1;z_1)]_{\mu_1',\mu_1}
[U_\varphi(a_2, \overline{a}_2; \overline{z_1})]_{\mu_2',\mu_2} U_\varphi(z_1, \overline{z}_1;0)}{U_\varphi(a_1,\overline{a}_1;0) U_{\varphi}(a_2, \overline{a}_2;0)}
|a_1, \overline{a_1};z_1,\mu_1 \rangle |a_2,\overline{a}_2;\overline{z}_1, \mu_2\rangle
\end{align}

It is straightforward to verify that in general the symmetry action on every basis state is unchanged by a natural isomorphism. Therefore, the symmetry action only depends on the equivalence class $[\varphi]$. This implies that if a braided autoquivalence $\varphi$ acts as the identity on the space of states on a torus while it acts non-trivially on surfaces of genus $g \geq 2$, then it must be a non-trivial soft symmetry.  

\subsection{2-group symmetry and $\mathcal{H}^3(\mathrm{Aut}(\mathcal{C}), \mathcal{A})$ class}

\begin{figure}[htbp]
    \centering
    \includegraphics[width=0.3\textwidth]{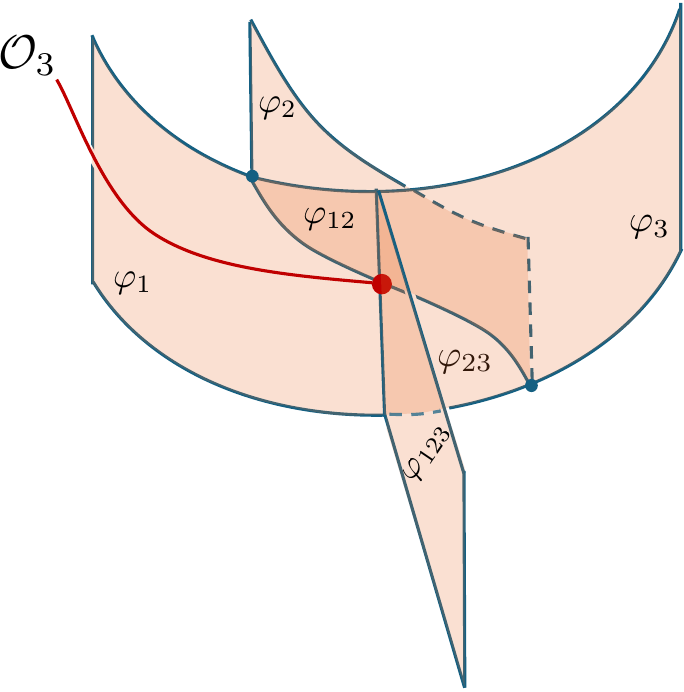}
    \caption{A point junction of the three symmetry defects emits an Abelian anyon $\mathcal{O}_3(\varphi_1,\varphi_2,\varphi_3)\in\mathcal{A}$. This represents the 2-group structure formed by the 0-form symmetries and the Abelian anyons.}
    \label{H3fig}
\end{figure}

The group of braided tensor autoequivalences defines an element of a 3rd cohomology group, $[\mathcal{O}_3] \in \mathcal{H}^3(\text{Aut}_{br}(\mathcal{C}), \mathcal{A})$, where $\mathcal{A}$ is a finite Abelian group defined by fusion of the Abelian anyons in $\mathcal{C}$ \cite{ENO2010,Barkeshli2019}. Formulas to compute $[\mathcal{O}_3]$ from the $U$-symbols were presented in \cite{Barkeshli2019}. Graphically, we can understand $[\mathcal{O}_3]$ as a failure of associativity for fusing the codimension-1 sheets. Consider three symmetries $[\varphi_1]$, $[\varphi_2]$, $[\varphi_3]$ $\in \text{Aut}_{br}(\mathcal{C})$ and the junctions formed by fusing the sheets as $\varphi_1 \circ (\varphi_2 \circ \varphi_3)$ compared with $(\varphi_1 \circ \varphi_2) \circ \varphi_3$. As we deform from the first order to the second, an Abelian anyon $O_3([\varphi_1], [\varphi_2], [\varphi_3]) \in \mathcal{A}$ is emitted from the resulting transition point in space-time. See Fig.~\ref{H3fig} for an illustration. We see how $[\mathcal{O}_3]$ can be viewed as an obstruction to associativity of the symmetry operations \cite{Barkeshli2019}.

The data $(\text{Aut}_{br}(\mathcal{C}), \mathcal{A}, [\mathcal{O}_3])$ defines a 2-group $\underline{\text{Aut}_{br}(\mathcal{C})}$ that acts on the UMTC $\mathcal{C}$ \cite{Barkeshli2019, Benini2018}. An example of a theory with non-trivial $[\mathcal{O}_3]$ was discussed in \cite{Barkeshli2019,fidkowski2015}, with generalizations where the symmetry is anti-unitary \cite{barkeshli2018}. 

\subsection{Symmetry fractionalization}

The existence of codimension-1 topological defects naturally leads to the possibility of codimension-2 junctions between different codimension-1 defects. In order to characterize the interaction between the anyon worldlines and these junctions, we need to introduce additional data, referred to as $\eta$-symbols \cite{Barkeshli2019}. These are depicted graphically as follows:
\begin{center}
\vspace{-5pt}
\raisebox{-.2\height}{\includegraphics[width=.65\linewidth]{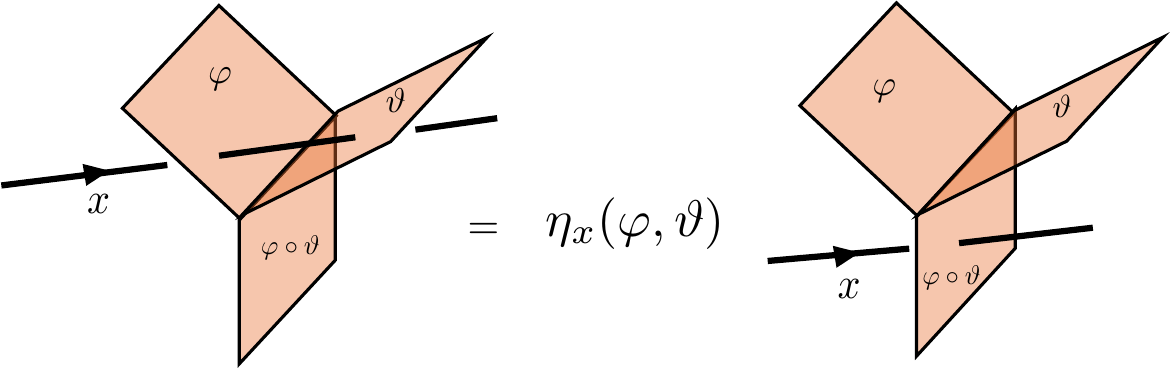}}
\end{center}
where $\varphi$, $\vartheta$ label two braided tensor autoequivalences.  

\begin{figure}[htbp]
    \centering
    \includegraphics[width=0.8\textwidth]{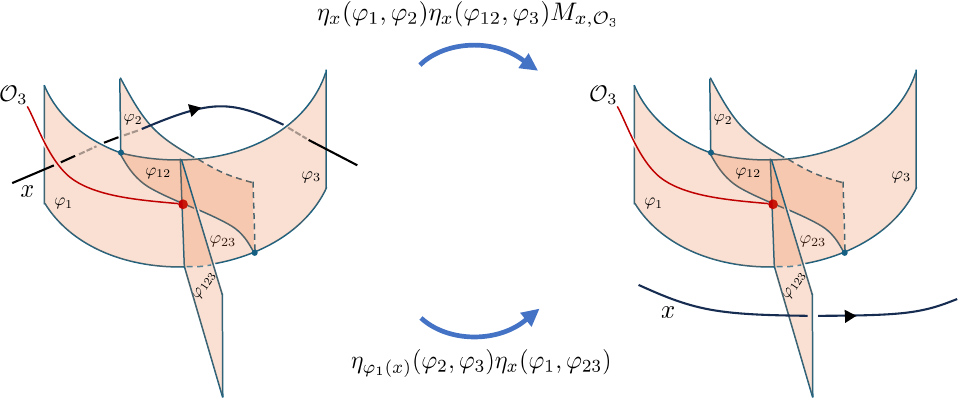}
    \caption{The anyon $x$ crossing through a pair of tri-junctions of symmetry defects. Initially $x$ is crossing the three defects $\varphi_1,\varphi_2,\varphi_3$, and finally $x$ crosses a single defect $\varphi_{123}$. There are two ways to arrive at the final configuration; passing $x$ along the top or bottom of the picture. The phase factor for each path is represented by the thick blue arrow. Equating the phase factors for consistency results in \eqref{eq:etaeta=Metaeta}.}
    \label{fig:H3crossing}
\end{figure}

By considering the fusion of 3 codimension-1 symmetry defects, Fig.~\ref{fig:H3crossing} illustrates how $\eta$ satisfies the equation
\begin{align}
    \eta_{\varphi_1(x)}(\varphi_2,\varphi_3) \eta_x(\varphi_1,\varphi_{23}) = M_{x, \mathcal{O}_3} \eta_x(\varphi_1,\varphi_2) \eta(\varphi_{12},\varphi_3),
    \label{eq:etaeta=Metaeta}
\end{align}
where $\varphi_{jk}$ denotes $\varphi_i\circ\varphi_j$, and $M_{xa}$ denotes the $U(1)$ phase resulting from a double braid of $x$ with an Abelian anyon $a \in \mathcal{A}$. If $[\mathcal{O}_3]$ is trivial in group cohomology $\mathcal{H}^3(\text{Aut}_{br}(\mathcal{C}), \mathcal{A})$ twisted by the action of $\text{Aut}_{br}(\mathcal{C})$ on $\mathcal{A}$, then there are inequivalent choices of the $\eta$ symbols, which form a torsor over the twisted cohomology $\mathcal{H}^2(\text{Aut}_{br}(\mathcal{C}), \mathcal{A})$. 

The $\eta$-symbols are referred to as the symmetry fractionalization data of the theory, because they specify how the anyons carry fractional quantum numbers under the symmetry group elements. For example, they can be used to characterize the fractional electric charge of anyons in the FQH effect or fractional spin of spinons in quantum spin liquids \cite{cheng2016lsm}. Assuming the $[\mathcal{O}_3]$ is trivial in $\mathcal{H}^3(\text{Aut}_{br}(\mathcal{C}), \mathcal{A})$, the $\eta$ symbols further determine an 't Hooft anomaly $[\mathcal{O}_4] \in \mathcal{H}^4(\text{Aut}_{br}(\mathcal{C}), U(1))$ \cite{ENO2010,Barkeshli2019,cui2016,Barkeshli2020Anomaly,Bulmash2020absolute}.   

\subsection{Global symmetry $G$ of quantum many-body system}

Consider a microscopic quantum many-body system with a global $0$-form symmetry group $G$. If the system forms a gapped phase with anyons described by a UMTC $\mathcal{C}$, then the possible symmetry-enriched topological phases will be partially characterized by a group homomorphism from the microscopic (UV) symmetry to the emergent (IR) symmetry,
\begin{align}
    [\rho]: G \rightarrow \text{Aut}_{br}(\mathcal{C}).
\end{align}
For each ${\bf g} \in G$, there will be a $U$-symbol
\begin{align}
U_{\bf g}(a,b;c) = U_{\rho({\bf g})}(a,b;c) . 
\end{align}
The group homomorphism $[\rho]$ can be pulled back to determine a cohomology class $[\rho]^*[\mathcal{O}_3] \in \mathcal{H}^3(G, \mathcal{A})$, which is an obstruction for the $\eta$ symbols of $G$ to satisfy associativity. Let us assume that $[\rho]^*[\mathcal{O}_3]$ is trivial. 

Symmetry fractionalization associated to junctions of codimension-1 symmetry defects of $G$ are given by
\begin{align}
    \eta_x({\bf g}, {\bf h}) = M_{x,{\bf t}({\bf g}, {\bf h})} \eta_x( \rho({\bf g}), \rho({\bf h})) ,
\end{align}
where $[{\bf t}] \in \mathcal{H}^2(G, \mathcal{A})$. Therefore, the possible choices of symmetry fractionalization for $G$ factor through those for Aut$_{br}(\mathcal{C})$, with an additional freedom specified by $[\bf t]$. The above data can be viewed as specifying a 2-homomorphism from $G$, viewed as a 2-group $\underline{G}$ with trivial $1$-form symmetry, to $\underline{\text{Aut}_{br}(\mathcal{C})}$:
\begin{align}
    \underline{\rho} : \underline{G} \rightarrow \underline{\text{Aut}_{br}(\mathcal{C})}.
\end{align}
If $[\rho]^*[\mathcal{O}_3]$ is non-trivial, then the theory is incompatible with an ordinary global $G$ symmetry. Instead we must consider a system with a non-trivial 2-group symmetry $\underline{G} = (G, \mathcal{A}, [\rho], [\rho]^*[\mathcal{O}_3])$. In this sense $[\rho]^*[\mathcal{O}_3]$ is an obstruction to having the symmetry be an ordinary group $G$.  

As mentioned above, the symmetry fractionalization data determines an element $[\mathcal{O}_4]\in \mathcal{H}^4(\text{Aut}_{br}(\mathcal{C}), U(1))$, which in the field theory literature is referred to as the 't Hooft anomaly. This induces an 't Hooft anomaly for $G$ via the pullback $[\mathcal{O}_{4,G}] = [\rho]^*[\mathcal{O}_4] \times \mathcal{O}_{4,\text{rel}}[{\bf t}]$, where $\mathcal{O}_{4,\text{rel}}[{\bf t}]$ is an additional contribution, referred to as a relative anomaly \cite{Barkeshli2020Anomaly}, determined by the choice $[\bf t]$, along with the $\eta$ and $U$ symbols for $\text{Aut}_{br}(\mathcal{C})$. General methods to compute these anomalies were discussed in \cite{ENO2010,Barkeshli2019,cui2016,Bulmash2020absolute,Barkeshli2020Anomaly}.

Note that the complete invertible symmetry of a UMTC is a 3-groupoid, $\underline{\underline{\text{Aut}_{br}(\mathcal{C})}}$.\footnote{This is referred to as a categorical 2-group in \cite{ENO2010}.} SETs with symmetry group $G$ are completely distinguished by a 3-homomorphism $\underline{\underline{\rho}}: \underline{\underline{G}} \rightarrow \underline{\underline{\text{Aut}_{br}(\mathcal{C})}}$. 
See \cite{ENO2010,Barkeshli2019,jones2020} for additional details. 

\section{Soft symmetry from gauged SPT defects}
\label{sec:soft}
\subsection{Review: gauged SPT defects}

Finite $G$ topological gauge theory is defined by the following partition function on a space-time manifold $M_{d+1}$:
\begin{align}
    Z(M_{d+1}) = \mathcal{N}^{-1} \sum_{[a] \in \mathrm{Hom}(\pi_1(M_{d+1}),G)/G }e^{i \int_{M_{d+1}} \omega_{d+1}}. 
\end{align}
Here $a$ is a flat $G$ gauge field and the sum is over all gauge inequivalent non-trivial flat $G$ gauge fields. $\omega_{d+1} \in Z^n(BG, U(1))$ is the Dijkgraaf-Witten twist \cite{dijkgraaf1990}, and we leave implicit the pullback to $Z^n(M_{d+1}, U(1))$ using $a$. In the following discussion, we assume that $\omega_{d+1}$ is trivial and we comment on non-trivial $\omega_{d+1}$ at the end. 

The above TQFT generally has finite invertible symmetries generated by operators called gauged SPT defects \cite{Barkeshli2023codim2,Barkeshli2024highergroup}. These symmetry defects are defined by the SPT topological response with support on the submanifolds where the defects are located. Namely, the topological defect in an $n$-dimensional submanifold $M_n$ embedded in $(d+1)$D spacetime is defined as
\begin{align}
    e^{i\int_{M_n}\omega(a)},
    \label{eq:gaugedSPT}
\end{align}
where $\omega\in Z^n(BG,U(1))$ is the topological response of a $G$-SPT phase in $n$ space-time dimensions \cite{dijkgraaf1990,chen2011}, and $a$ is the $G$ gauge field. From the perspective of the partition function, the above operator is a symmetry because it can be inserted in the sum in the partition function and the result is independent of smooth deformations of the location of $M_n \subset M_{d+1}$, as long as they are away from other defects. From the lattice model perspective that we introduce in the subsequent section, the action of this gauged SPT operator can be understood as pumping the SPT phase through the space by the action of a finite depth local unitary quantum circuit. 

\begin{figure}[t]
    \centering
    \includegraphics[width=0.9\textwidth]{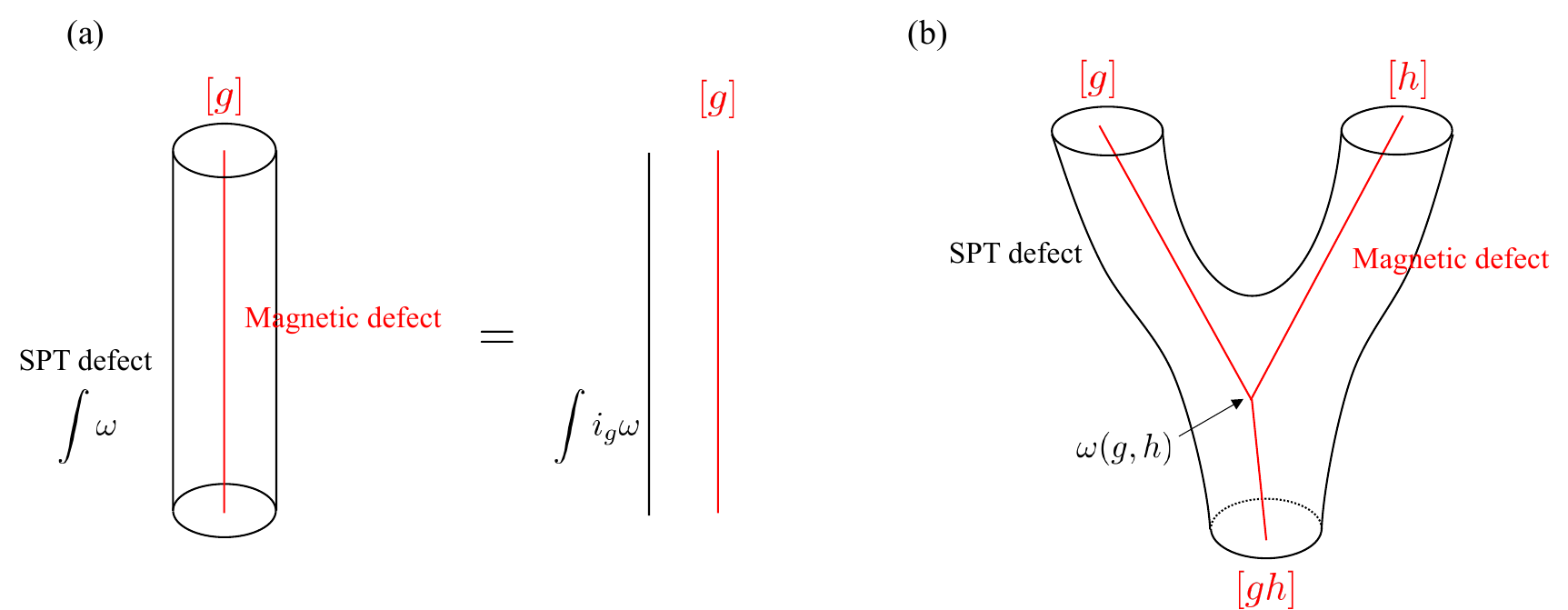}
    \caption{The action of a gauged SPT defect on magnetic defects. (a): The 0-form symmetry generated by the gauged SPT defect acts on the magnetic defect by attaching the $(d-1)$D gauged SPT defect to the magnetic defect. This can be understood as the twisted compactification of SPT with respect to the circle with holonomy $g\in G$.
    (b): The 0-form gauged SPT defect in $(d+1)$D spacetime acts on a point junction of the $(d-1)$ magnetic defects by a phase factor given by the SPT cocycle.  The figure shows the case of $d=2$, where the (1+1)D SPT defect acts on the junction of two magnetic defects by a 2-cocycle $\omega(g,h)$.}
    \label{fig:SPTaction}
\end{figure}

The gauged SPT defects generally act on the codimension-2 magnetic defects of the $G$ gauge theory that correspond to fluxes. For instance, the 0-form symmetry \eqref{eq:gaugedSPT} with $n=d$ acts on the magnetic fluxes by permuting codimension-2 topological defects. When the magnetic flux $V_{[g]}$ is labeled by the conjugacy class $[g]$ with $g\in G$, the gauged SPT defect acts by the automorphism of symmetry defects as
\begin{align}
    V_{[g]}(M_{d-1}) \to V_{[g]}(M_{d-1}) = V_{[g]}(M_{d-1}) \times e^{i\int_{M_{d-1}}i_g\omega}~.
\end{align}
Here $i_g$ is the slant product defined through the circle compactification of the SPT response, $i_g\omega:= \int_{S^1_{g}}\omega$. Note that when $G$ is non-Abelian and $\omega\in Z^n(BG,U(1)),$ $i_g\omega\in Z^{n-1}(BZ(g),U(1))$, where $Z(g)\subset G$ is the centralizer of $G$. This reflects that the flat $G$ gauge field on $M_{n-1}\times S^1_g$ must have the holonomy in $Z(g)$ along any cycle of $M_{n-1}$. 
In (2+1)D gauge theories with $d=2$, this action is understood as the permutation of anyons. See Fig.~\ref{fig:SPTaction} (a).

One can use a specific gauged SPT defect to define a soft auto-equivalence of $G$ gauge theory in (2+1)D. When $\int\omega$ is trivial on a torus, the slant product of the cocycle $\omega\in Z^2(BG,U(1))$ is trivial in cohomology. Therefore a gauged SPT defect with such an $\omega$ does not permute anyons and defines a soft auto-equivalence. Such a soft auto-equivalence was first found by Davydov in \cite{Davydov_2014}.

The gauged SPT defects also act on the junctions of the magnetic defects. For instance, in (2+1)D gauge theories suppose we have an operator corresponding to a network of magnetic fluxes. Then the gauged SPT defect acts non-trivially on the junctions of two magnetic fluxes $[g],[h]\in Cl(G)$ into $[gh]$, where $Cl(G)$ denotes the set of conjugacy classes of $G$. The 0-form symmetry generated by the (1+1)D gauged SPT defect acts on this junction by a phase factor $\exp(i\omega(g,h))$, since the SPT response evaluates nontrivially at this junction of $G$ defects. This implies that the symmetry operator acts on the junction of magnetic fluxes by a phase $\mathcal{U}_\omega$ given by
\begin{align}
    \mathcal{U}_\omega([g],[h];[gh]) = e^{i\omega(g,h)}~, 
\end{align}
which is valid when $\omega(g,h)$ is invariant under overall conjugation by group elements, $\omega(g,h)=\omega(kgk^{-1},khk^{-1})$, so that the above action does not depend on the choices of the representatives in the conjugacy classes. See Fig.~\ref{fig:SPTaction} (b).

When the gauged SPT defect does not permute anyons, the above phase factor $\mathcal{U}_\omega$ leaves the $F$ symbols invariant, which follows from the cocycle condition of satisfied by $\omega$.
When the cocycle $\omega\in Z^{2}(BG,U(1))$ is deliberately chosen such that the above phases $\mathcal{U}_\omega$ also leaves the $R$ symbols invariant, we identify the phase $\mathcal{U}_\omega$ with the $U$ symbol of the soft braided auto-equivalence,
\begin{align}
    U([g],[h];[gh]) = \mathcal{U}_\omega([g],[h];[gh])~.
\end{align}
We will verify this formula explicitly in Sec.~\ref{sec:double} in the context of a lattice model and a particular example of a gauged SPT.  Meanwhile, the junction of the SPT defects acts trivially on the magnetic fluxes, so we have $\eta=1$. 

In general, the gauged SPT defects with distinct dimensions combined with the magnetic defects of the $G$ gauge theory form the algebraic structure of a higher fusion category. See \cite{Barkeshli2024highergroup} for detailed discussions.

 Below, we will discuss a physical perspective of such soft auto-equivalences in detail.

\subsection{Soft autoequivalence of finite gauge theory}
\label{sec:softexample_gen}

Suppose that the braided fusion category is given by a (2+1)D finite gauge theory $\mathcal{C}=Z(\text{Vec}(G))$, where $Z(\text{Vec}(G))$ denotes the Drinfeld center of $\text{Vec}(G)$.
Its simple anyons are labeled by a pair
\begin{align}
    ([g],\pi)~,
    \label{eq:anyonlabel}
\end{align}
where $[g]\in Cl(G)$ is a conjugacy class of $G$ that specifies the magnetic flux, and $\pi\in \text{Rep}(Z(G))$ is the irreducible representation that corresponds to the electric charge.

Then, a subgroup of the soft braided auto-equivalence $\mathrm{Aut}_{\mathrm{sf}}(\mathcal{C})$ has the structure of
\begin{align}
    \mathrm{Out}_{\mathrm{sf}}(G)\ltimes H^2_{\mathrm{sf}}(BG,U(1))\subset \mathrm{Aut}_{\mathrm{sf}}(\mathcal{C})~.
\end{align}
\begin{itemize}
    \item  $\mathrm{Out}_{\mathrm{sf}}(G)$ is a subgroup of the outer automorphism $\mathrm{Out}(G)$ that preserves the conjugacy class of any commuting pair $g,h\in G$. In \cite{Davydov_2014}, such outer automorphisms are referred to as doubly class preserving outer automorphisms. 
Given that the $G$ gauge field on a torus is labeled by the commuting pair of groups, $\mathrm{Out}_{\mathrm{sf}}(G)$ corresponds to the symmetry that trivially acts on the torus Hilbert space. Hence this subgroup does not permute anyons. See \cite{Neumann_1981, Szechtman2003, Davydov_2014} for examples of such outer automorphisms.
\item $H^2_{\mathrm{sf}}(BG,U(1))$ is a subgroup of group cohomology $H^2(BG,U(1))$, given by a set of $\omega\in H^2(BG,U(1))$ whose integral on a torus with any flat $G$ gauge field becomes trivial, $\int_{T^2}\omega=0 \mod 2\pi \mathbb{Z}$.
The symmetry $H^2(BG,U(1))$ is generated by gauged SPT defects, and $H^2_{\mathrm{sf}}(BG,U(1))$ corresponds to the ones acting trivially on the torus Hilbert space.
\end{itemize}

Let us provide an example of finite gauge theory which has the soft symmetry.
We consider the untwisted (2+1)D gauge theory with gauge group $G=\gsf$, where $\gsf$ is a group with order 128, defined by the following group algebra involving order-2 elements $p,q,r_1,r_2, x, y_1, y_2$~\cite{Pollmann2012detection, Davydov_2014},
\begin{align}
    \begin{split}
        pqp^{-1}q^{-1} &= x, \\
        r_1r_2 r_1^{-1}r_2^{-1} &= x, \\
        pr_1 p^{-1} r_1^{-1} &= y_1, \quad p r_2 p^{-1} r_2^{-1} = y_2, \\ 
        qr_1 q^{-1} r_1^{-1} &= qr_2 q^{-1} r_2^{-1} = 1.
    \end{split}
    \label{eq:multiplication}
\end{align}
This group $\gsf$ is regarded as a central extension $(\Z_2)^3\to \gsf \to (\Z_{2})^4$, where $(\Z_2)^3$ is generated by $\{x, y_1, y_2\}$ and $(\Z_{2})^4$ is generated by $\{p,q,r_1,r_2\}$. 
Let $n_p, n_q\in Z^1(\gsf,\Z_2)$ be the $\Z_2$ cocycle such that $n_p(g)$ (resp.~$n_q(g)$) for $g\in \gsf$ gives the mod 2 number of the element $p$ (resp.~$q$) that appears in the expression of $g$ in terms of the multiplication of generators $p,q,r_1,r_2, x, y_1, y_2$.

We then consider the codimension-1 gauged SPT defect generated by the (1+1)D $\gsf$-SPT phase with the action given by a 2-cocycle
\begin{align}
    \omega_{pq}=\pi (n_p\cup n_q)\in Z^2(B\gsf,U(1))
\end{align}
where the factor $\pi$ originates from the $2\pi$ periodicity of $U(1)$. This generates the $\Z_2$ 0-form symmetry of the (2+1)D $\gsf$ gauge theory. With the above choice of cocycle representative of $H^2(B\gsf,U(1))$, the action on the anyon junctions are given by the phase factor
\begin{align}
    \mathcal{U}_{pq}(([g_1],\pi_1),([g_2],\pi_2);([g_3],\pi_3)) = e^{i\omega_{pq}(g_1,g_2)},
    \label{eq:Usymboloriginal}
\end{align}
where we label the anyon by the conjugacy class $[g_j]$ with $g_j\in \gsf$ and the irreducible representation of the centralizer group $\pi_j\in \mathrm{Rep}(Z(g_j))$. 
Note that the above phase $\mathcal{U}_{pq}$ does not depend on the choice of representatives of the conjugacy class.
We will later see that this symmetry does not permute anyons. 
As required by \eqref{Finvariance}, the symmetry action leaves the $F$ symbol invariant, which is seen from the cocycle condition of group cohomology satisfied by $\mathcal{U}_{pq}$. 

However, the above phase $\mathcal{U}_{pq}$ presented in \eqref{eq:Usymboloriginal} does not leave the $R$ symbols invariant. It transforms the $R$ symbol into
\begin{align}
    R^{([g_1],\pi_1),([g_2],\pi_2)}_{([g_3],\pi_3)} \to \exp\left(i(\omega_{pq}(g_1,g_2)-\omega_{pq}(g_2,g_1))\right) R^{([g_1],\pi_1),([g_2],\pi_2)}_{([g_3],\pi_3)}
\end{align}
In particular, when $g_1=p,g_2=q$ the $R$ symbol is transformed into $-R$, so it does not satisfy \eqref{Rinvariance}. 
Meanwhile, one can find another representative $\omega_{\text{sf}}\in Z^2(B\gsf,U(1))$ in the same cohomology class as $\omega_{pq}$, i.e., $[\omega_{\text{sf}}]=[\omega_{pq}]\in H^2(B\gsf,U(1))$ and symmetric as a function of $g,h$; $\omega_{\text{sf}}(g,h)=\omega_{\text{sf}}(h,g)$ for any $g,h\in\gsf$.
We will explicitly construct such a cocycle representative in Appendix \ref{app:cocycle}.
With this new choice of 2-cocycle, the action on fusion vertices are given by
\begin{align}
    U_{\mathrm{sf}}(([g_1],\pi_1),([g_2],\pi_2);([g_3],\pi_3)) = e^{i\omega_{\mathrm{sf}}(g_1,g_2)},
    \label{eq:Usymbolnew}
\end{align}
and this phase action leaves both $F$, $R$ symbols invariant, being consistent with \eqref{Finvariance}, \eqref{Rinvariance}. 
From now, we will use $U_{\text{sf}}$ as the data of the braided tensor auto-equivalence.
This symmetry is regarded as the vertex basis gauge transformation with the matrices given by the phases $\Gamma=U_{\text{sf}}$. The symmetry does not permute anyons, while as demonstrated below, this vertex gauge transformation is not a natural isomorphism.
In other words, this $\Z_2$ symmetry is a soft symmetry.
Below, let us summarize the notable properties of this $\Z_2$ symmetry:
\begin{itemize}
    \item The $\Z_2$ symmetry does not induce the permutation of anyons. To see this, recall that the SPT defects $\omega_2$ acts on the magnetic flux $[g]$ by attaching an electric charge given by the slant product $i_g\omega_2\in \mathrm{Rep}(Z(g))$. The slant product is related to the torus partition function through $i_g\omega_2(h)=Z_{\text{SPT}}(T^2(g,h))$, where $T^2(g,h)$ denotes a torus with the holonomy $g,h\in\gsf$ on each cycle. 
    For the cocycle $\omega_{\text{sf}}$, the slant product is trivial since the torus partition function with any flat $\gsf$ gauge field becomes trivial, $Z_{\text{SPT}}=1$. For instance, if $n_p(g)=1$ on a torus $T^2(g,h)$, $h$ needs to satisfy $n_q(h)=0$ due to $ghg^{-1}h^{-1}=1$, so the integral of $\omega_{\text{sf}}$ (which is identical to the integral of $\omega_{pq}$) vanishes.
    This leads to the trivial permutation action by the symmetry.

    \item  This $\Z_2$ soft symmetry is faithful in the sense that it acts by a non-trivial operator on the Hilbert space supported on a genus $g \geq 2$ surface. 
    For instance, the action on the genus 2 surface with the anyon diagrams $a_1=[p], a_2=[r_1], z= [x]$ in Fig.~\ref{fig:genusg} is given by
\begin{align}
\begin{split}
\varphi_{\text{sf}}(|[p],[p];[x]\rangle |[r_1],[r_1];[x]\rangle) &= 
\frac{ U_{\text{sf}}([p],[p];[x])
U_{\text{sf}}([r_1],[r_1];[x]) U_{\text{sf}}([x],[x];0)}{U_{\text{sf}}([p],[p];0) U_{\text{sf}}([r_1],[r_1];0)}
|[p],[p];[x]\rangle |[r_1],[r_1];[x]\rangle \\
&= \frac{\exp\left(i(\omega_{\text{sf}}(p,px)+\omega_{\text{sf}}(r_1,r_1x)+\omega_{\text{sf}}(x,x))\right)}{\exp\left(i(\omega_{\text{sf}}(p,p)+\omega_{\text{sf}}(r_1,r_1)\right)}|[p],[p];[x]\rangle |[r_1],[r_1];[x]\rangle \\
&= - |[p],[p];[x]\rangle |[r_1],[r_1];[x]\rangle
\end{split}
\end{align}   
which can be explicitly computed by using $\omega_{\text{sf}}$ presented in Appendix \ref{app:cocycle}. Since the above eigenvalue is invariant under the natural isomorphisms, this nontrivial invariant $(-1)$ implies that this soft $\Z_2$ symmetry corresponds to a nontrivial braided tensor auto-equivalence which cannot be given by natural isomorphisms.
    In Sec.~\ref{sec:double}, this $\Z_2$ symmetry will be  realized in a non-Pauli stabilizer model of qubits. Therefore, this $\Z_2$ symmetry gives a non-trivial logical gate of the stabilizer code that does not permute anyons.

    \item  We have $\eta=1$, meaning that the anyons do not carry fractional charge under the $\Z_2$ symmetry. 
    
    \item This $\Z_2$ soft symmetry can be gauged. The $\Z_2$ gauge field $A$ for the soft symmetry is introduced by the action $\int A \cup \omega_{\text{sf}}$, so the resulting gauge theory is the $\gsf\times \Z_2$ gauge theory with the Dijkgraaf-Witten twist given by $A\cup \omega_{\text{sf}}\in H^3(B(\gsf\times\Z_2),U(1))$. Note that this theory is distinct from the untwisted $\gsf\times\Z_2$ gauge theory. For instance, in the twisted gauge theory the magnetic defect of the $\Z_2$ soft symmetry 
    becomes non-invertible. This is because it corresponds to the endpoint of the (1+1)D gauged SPT, which carries the projective representation of $\gsf$ symmetry characterized by $\omega_{\text{sf}}\in H^2(B\gsf,U(1))$.
    \end{itemize}

While we discussed the soft symmetry in the untwisted $\gsf$ gauge theory, this soft $\Z_2$ symmetry is present in $\gsf$ gauge theory with a Dijkgraaf-Witten twist  $\omega_3$ \cite{dijkgraaf1990} as well, as long as the 2-cocycle $\omega_{\text{sf}}$ is not expressed as a slant product of $\omega_{\text{sf}}=i_g\omega_3$ with some $g\in\gsf$. When $\omega_{\text{sf}}$ is the slant product, 
this SPT defect is bounded by the topological line operator of the magnetic flux. Concretely, suppose $\tilde a$ is the dual gauge field for $g\in\gsf$, meaning $\exp(i\int \tilde{a})$ gives the line operator for the magnetic defect. We then have $d\tilde a = i_g\omega_3 \mod 2\pi \mathbb{Z}$.\footnote{More precisely, we have $d\tilde a = A^*(i_g\omega_3)$ where $A: M_3\to B\gsf$ is the background $\gsf$ gauge field on the spacetime $M_3$.} This means that the surface operator $\exp(i\int \omega_{\text{sf}})$ is bounded by the $g$ magnetic defect when $i_g\omega_3 = \omega_{\text{sf}} \mod 2\pi \mathbb{Z}$. 
Since $\omega_{\text{sf}}$ is a coboundary, $\exp(i\int \omega_{\text{sf}})$ does not generate a faithful 0-form symmetry~\cite{Barkeshli2024highergroup}.

Also, we note that (2+1)D finite gauge theories can have a ``softer'' 0-form symmetry, which acts trivially on the Hilbert space of a genus $< g$ surface, but acts non-trivially on the genus $g$ surface. Such a global symmetry can be generated by the gauged SPT defects, where the SPT can only be detected by its partition function on higher genus surfaces. Such examples are explicitly constructed in Appendix \ref{app:highergenus}. It would be interesting to see if such softer symmetry can also arise from the outer automorphisms of gauge groups.

\section{Lattice model}
\label{sec:double}

\subsection{Quantum double model}

The above $\Z_2$ symmetry action on the $\gsf$ gauge theory can be explicitly demonstrated on the quantum double model~\cite{Kitaev2003double}. 
The lattice model can be defined on generic triangulations of a 2d surface. For simplicity, let us consider a square lattice with seven qubits at each edge. One can describe the (2+1)D quantum double model with the gauge group $\gsf$ in this setup.

The local Hilbert space on each edge has $|\gsf|=2^7$ dimensions, whose basis states $\{\ket{g}\}$ are labeled by group elements $g\in \gsf$. The Hamiltonian is given by
\begin{align}
    H_{\gsf} =  -\sum_v A_v - \sum_f B_f 
    \label{eq:double}
\end{align}
with each term given by
\begin{align}
    A_v = \frac{1}{|\gsf|}\sum_{g\in \gsf}\overrightarrow{X}_{N(v)}^g\overrightarrow{X}_{E(v)}^g\overleftarrow{X}_{W(v)}^{g^{-1}}\overleftarrow{X}_{S(v)}^{g^{-1}}, \quad B_f = \delta_{g_{01}g_{13}g^{-1}_{23}g^{-1}_{02},0}~.
\end{align}
Here we defined the Pauli $X$ like operators as
\begin{align}
    \overrightarrow{X}^g\ket{h} = \ket{gh}, \quad \overleftarrow{X}^{g^{-1}}\ket{h} = \ket{hg^{-1}}~.
\end{align}

\begin{figure}[t]
    \centering
    \includegraphics[width=0.5\textwidth]{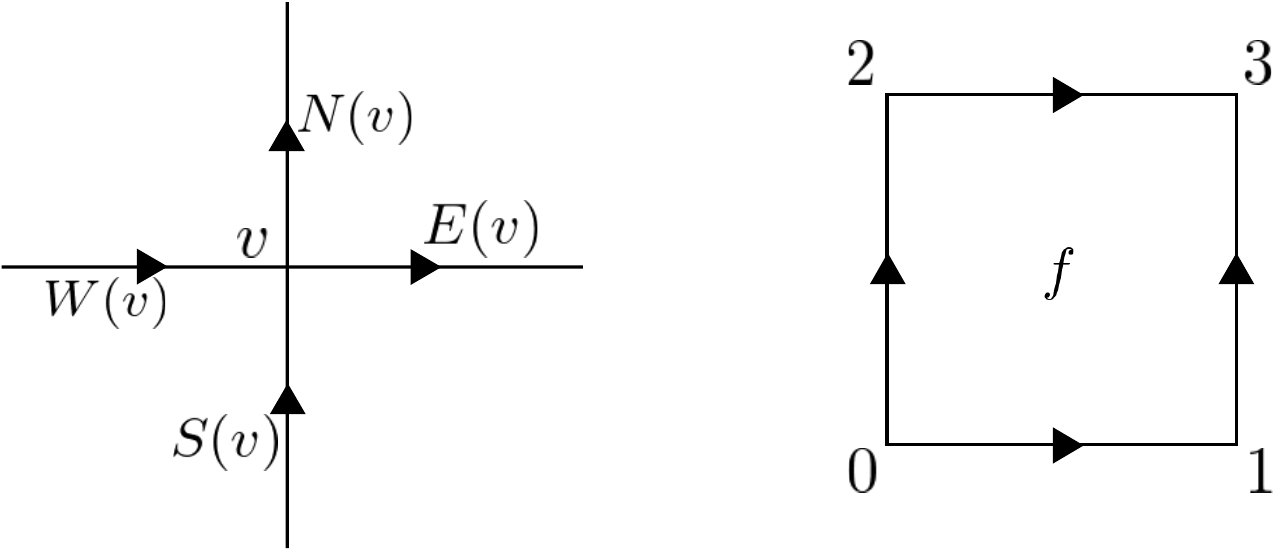}
    \caption{The edges nearby a vertex $v$ and face $f$.}
    \label{fig:doublemodel}
\end{figure}

We label $g\in \gsf$ by a pair $g =((p,q,r_1,r_2),(x,y_1,y_2))$ with $(p,q,r_1,r_2)\in \Z_2^4, (x,y_1,y_2)\in \Z_2^3$, satisfying the group multiplication law \eqref{eq:multiplication}. 
When we express the local Hilbert space with seven qubits, the above labels $((p,q,r_1,r_2),(x,y_1,y_2))$ are identified as the eigenvalues of the Pauli $Z$ operators as
\begin{align}
    Z^{(j)}=(-1)^{\alpha_j} \quad (1\le j\le 7)\end{align}
where $\{\alpha_j\}=(p,q,r_1,r_2,x,y_1,y_2)$. The above quantum double model is regarded as a non-Pauli (Clifford) stabilizer model of qubits.

The symmetry $\mathcal{R}_{\Z_2}$ is then represented as the product of $CZ$ operators
\begin{align}
    \mathcal{R}_{\Z_2} = \prod_{f=(0123)} C(Z^{(1)}_{01},Z^{(2)}_{13}) C(Z^{(1)}_{02},Z^{(2)}_{23}).
\end{align}
where the subscripts $01,...$ label the edges of a face $(0123)$, see Fig.~\ref{fig:doublemodel}. $C(Z,Z')$ means the controlled-Z operator involving two Pauli $Z$ operators $Z,Z'$. This operator on each face $f$ is regarded as evaluating $\omega_{pq}=\pi (n_p\cup n_q)$ on a face,
\begin{align}
    \mathcal{R}_{\Z_2} = \prod_{f=(0123)} e^{i\omega_{pq}(0123)}
    \label{eq:productomega}
\end{align}
since $\omega_{pq}(0123) = \pi (n_p(01)n_q(13) + n_p(02)n_q(23))$.\footnote{See e.g., \cite{Chen2023cup} for the details of cup product on square and hypercubic lattices.} Therefore, this operator corresponds to the action of the gauged SPT operator $e^{i\int\omega_{pq}}$ acting on the flat $G_{\text{sf}}$ gauge fields on a 2d space. $\mathcal{R}_{\mathbb{Z}_2}$ remains unchanged on a closed surface if we use a different cocycle representative for $\omega$. In particular, since we have $[\omega_{pq}]=[\omega_{\text{sf}}]$ in cohomology $H^2(B\gsf,U(1))$, the above $\mathcal{R}_{\Z_2}$ can also be expressed as
    $\mathcal{R}_{\Z_2} = \prod_{f=(0123)} e^{i\omega_{\text{sf}}(0123)}$, which corresponds to an alternative way of decomposing the finite depth circuit $\mathcal{R}_{\mathbb{Z}_2}$ into local unitaries.

One can define the operator $\mathcal{R}_{\Z_2}$ in the form of \eqref{eq:productomega} on generic triangulation in 2d equipped with branching structure, where the operator is expressed as
\begin{align}
    \mathcal{R}_{\Z_2} = \prod_{\Delta = (012)}(-1)^{n_p(01)n_q(12)}~,
\end{align}
where the product is over 2-simplices of the space.

The operator $\mathcal{R}_{\Z_2}$ commutes with the Hamiltonian within the Hilbert space satisfying $B_f=1$ for any faces. First of all, it is obvious that $\mathcal{R}_{\Z_2}$ commutes with $B_f$ itself.
Then, the action of $\mathcal{R}_{\Z_2}$ on the $X$ stabilizers are given by
\begin{align}
    \mathcal{R}_{\Z_2} (\overrightarrow{X}_{N(v)}^g\overrightarrow{X}_{E(v)}^g\overleftarrow{X}_{W(v)}^{g^{-1}}\overleftarrow{X}_{S(v)}^{g^{-1}}) \mathcal{R}_{\Z_2}^\dagger = (\overrightarrow{X}_{N(v)}^g\overrightarrow{X}_{E(v)}^g\overleftarrow{X}_{W(v)}^{g^{-1}}\overleftarrow{X}_{S(v)}^{g^{-1}}) \prod_{e\subset f_{\text{ws}}} Z_e^{(1)} \prod_{e\subset f_{\text{ne}}} Z_e^{(2)} 
    \end{align}
where $f_{\text{ws}}$ (resp.~$f_{\text{ne}}$) is the face bounded by two edges $W(v),S(v)$ (resp.~$N(v),E(v)$). The above product of Pauli $Z$ operators becomes trivial when $B_f=1$ is satisfied, so $\mathcal{R}_{\Z_2}$ commutes with $X$ terms of the Hamiltonian within this Hilbert space. This implies that $\mathcal{R}_{\Z_2}$ generates an emergent $\Z_2$ symmetry that preserves the ground state subspace.

\subsection{Trivial permutation action on anyons}

The anyons of the quantum double model are created and moved using ribbon operators \cite{Kitaev2003double, Bombin2008double}. Here we demonstrate that the $\Z_2$ symmetry does not permute anyons by explicitly checking the symmetry action on closed ribbon operators.

\begin{figure}[t]
    \centering
    \includegraphics[width=0.7\textwidth]{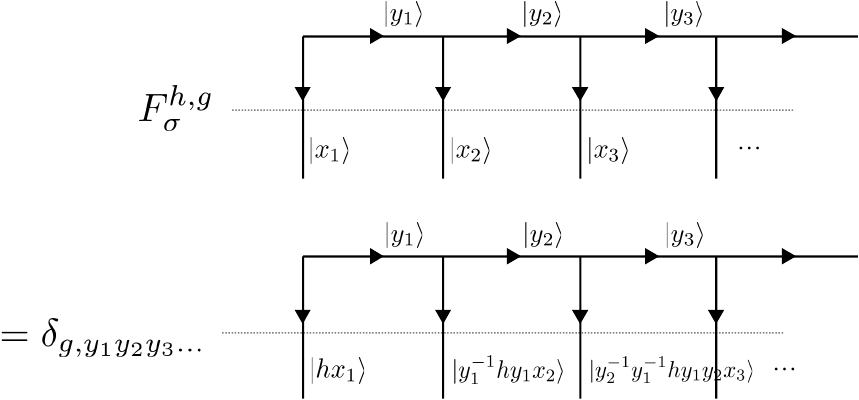}
    \caption{The ribbon operator of the quantum double model. The operator $F^{h,g}$ generates the gauge transformation $h\in G$ along the dual edges cutting the ribbon, associated with the projection onto the holonomy $g\in G$ along the ribbon.
    This can be defined on a generic curve of the dual lattice, see \cite{Kitaev2003double, Bombin2008double} for details.}
    \label{fig:ribbon}
\end{figure}

Suppose we pick a conjugacy class $C\in Cl(G)$ where $Cl(G)$ denotes the set of conjugacy classes of $G$.  We fix a representative $r_C\in C$, and define $Q_C=\{s_i\}$ as a set of group elements with $|Q_C|=|C|$, such that for $C=\{c_i\}$, $s_i$ is defined as an element satisfying $c_i = s_i r_C \overline{s}_i$. Also, let us pick a conjugacy class $D\in  Cl(Z(r_C))$. 
The ribbon operator on a closed ribbon $\sigma$ can be labeled by a pair $\{D,C\}$ as~\cite{Bombin2008double}
\begin{align}
    K_\sigma^{DC} = \sum_{s\in Q_C} \sum_{k\in D} F_\sigma^{s r_C \overline{s},s k \overline{s}}  
    \label{eq:closedribbonDC}
\end{align}
which uses the building block of ribbon operators $F^{h,g}$ shown in Fig.~\ref{fig:ribbon}. This operator $F^{h,g}$ generates the $h\in G$ gauge transformation along the dual edges cutting the ribbon, and associated with the projection onto the holonomy $g\in G$ along the ribbon.

Then, using an irrep $R\in \text{Rep}(Z(r_C))$ one can express the ribbon operators in the standard basis
\begin{align}
    K_\sigma^{RC} = \frac{\text{dim}(R)}{|Z(r_C)|}\sum_{D\in  Cl(Z(r_C))} \overline{\chi}_R(D)K_\sigma^{DC}~,
\end{align}
where $C$ labels the magnetic flux, and $R$ labels the electric charge attached to it. This corresponds to the closed string of the anyon labeled by a pair $(C,R)$ as introduced in \eqref{eq:anyonlabel}.

One can check that the symmetry operator $\mathcal{R}_{\Z_2}$ preserves the above closed ribbons $K_\sigma^{RC}$. 
For instance, let us take a conjugacy class $C=[p]$ where $p,q\in\gsf$ is introduced in \eqref{eq:multiplication}. In that case, the $\Z_2$ symmetry acts on $K^{R,[p]}_\sigma$ by attaching the string of Pauli $Z^{(2)}$ operators parallel to the ribbon $\sigma$,
\begin{align}
    \mathcal{R}_{\Z_2}K^{R,[p]}_\sigma \mathcal{R}^\dagger_{\Z_2} = \left(\prod_{e\in \sigma} Z_e^{(2)}\right) K^{R,[p]}_\sigma
\end{align}
The string of $Z^{(2)}$ measures the holonomy of $q\in \gsf$ along the ribbon $\sigma$. Here, each ribbon $K^{D,[p]}_\sigma$ involves a projection of holonomy onto some group element $sk\overline{s}$ with $k\in Z(p)$. Since $sk\overline{s}$ must contain even number of $q\in\gsf$ in its expression (i.e., $n_q(sk\overline{s})=0$ mod 2), the product of $Z^{(2)}$ must evaluate trivially after such projection. Therefore we have
\begin{align}
    \mathcal{R}_{\Z_2}K^{R,[p]}_\sigma \mathcal{R}^\dagger_{\Z_2} =  K^{R,[p]}_\sigma
\end{align}
hence it preserves the closed ribbon operator. Similar logic is also valid for other choices of $C\in Cl(\gsf)$. 

\subsection{Symmetry action on anyon junctions}

\begin{figure}[t]
    \centering
    \includegraphics[width=\textwidth]{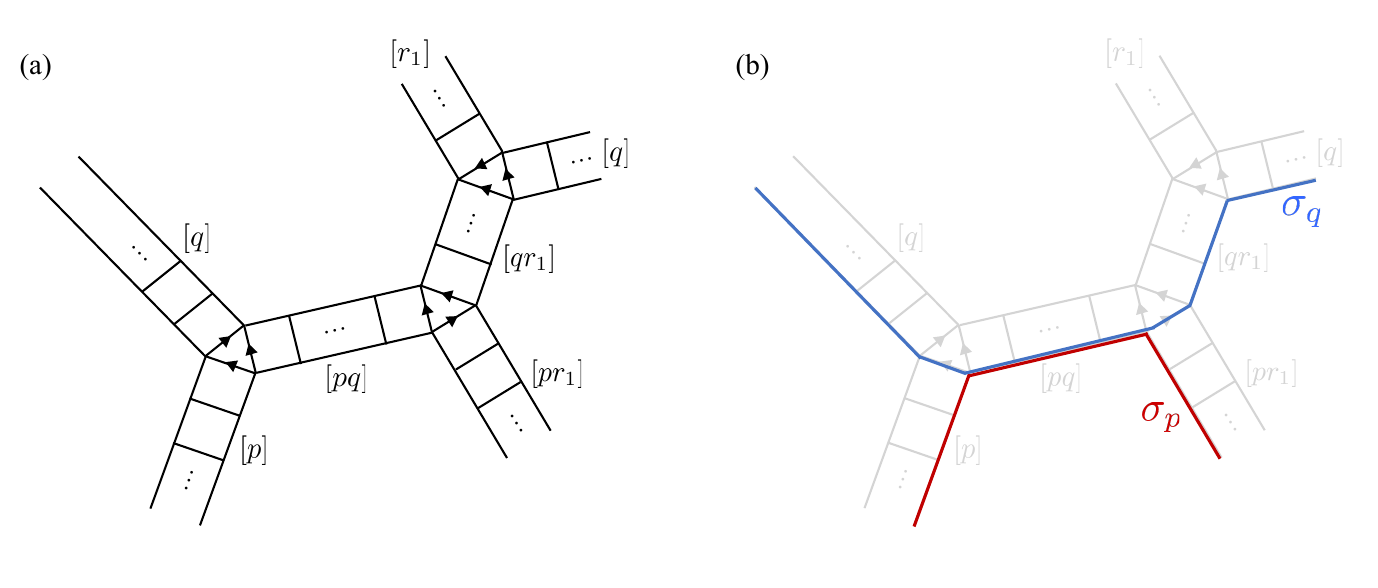}
    \caption{The network of the ribbon operators, with the ribbons connected by the tri-junctions. (a): an example of the network where each ribbon carries magnetic flux labeled by the conjugacy class of $\gsf$. (b): Closed curves $\sigma_p$ (resp.~$\sigma_q$) are defined by picking up the ribbons carrying the magnetic flux with $n_p(g)=1$ (resp.~$n_q(g)=1$.)}
    \label{fig:network}
\end{figure}

Though the $\Z_2$ symmetry does not permute anyons, the $\Z_2$ symmetry acts on the junction of anyon line operators. Let us consider the closed network of the ribbon operators that involves the tri-junctions of the ribbons. This can be regarded as a network of magnetic fluxes, where we have the operator  $\overrightarrow{X}^g$ along each dual edge cutting the network. Three ribbon operators carrying the magnetic fluxes $g,h,gh\in \gsf$ can meet at a junction which is a single triangular face. See Fig.~\ref{fig:network} (a) for an example.

Let us pick a set of dual edges on which the magnetic flux $g\in \gsf$ has an odd number of $p\in \gsf$ in its expression, i.e., $n_p(g)=1$ mod 2. This gives a set of closed curves $\sigma_p$ contained in the network. Let us similarly define a set of closed curves $\sigma_q$ for $q\in\gsf$. See Fig.~\ref{fig:network} (b). Then, the ribbon operator network $K$ has to include the projector onto the holonomy satisfying $n_q(g)=0$ along each closed curve of $\sigma_p$, and  holonomy with $n_p(g)=0$ along each closed curve of $\sigma_q$. Similar to the closed ribbon operators introduced in \eqref{eq:closedribbonDC}, such projections are necessary to make the gauge transformations along the closed network well-defined.
With this in mind, the $\Z_2$ action on the ribbon network is given in the form of
\begin{align}
    \mathcal{R}_{\Z_2} K \mathcal{R}^\dagger_{\Z_2} \propto \left(\prod_{e\in \sigma_p} Z^{(2)}_e \prod_{e\in \sigma_q} Z^{(1)}_e \right) K = K,
    \label{eq:action on junctions}
\end{align}
where we used the fact that $K$ has the projectors onto trivial holonomy in the last equation.
The $\propto$ in the first equation means that the equation is valid only up to phase; we will shortly see that the phase is carried by each anyon junction in the network, and this phase corresponds to the symmetry action on the anyon junctions.

To fix the phase ambiguity of \eqref{eq:action on junctions}, let us consider the junction of magnetic fluxes $g,h\in \gsf$ into $gh$ at a triangle $(012)$. 
Most generally, the edge $(01)$ supports an operator $\overrightarrow{X}^{sg\overline{s}}$, edge $(12)$ supports 
$\overrightarrow{X}^{th\overline{t}}$, and edge $(02)$ supports $\overrightarrow{X}^{ugh\overline{u}}$ for some $s,t,u\in\gsf$. Focusing on the operators nearby this junction, the action of $\mathcal{R}_{\Z_2}$ can be expressed as
\begin{align}
\begin{split}
    \mathcal{R}_{\Z_2} K \mathcal{R}^\dagger_{\Z_2} &= \dots(Z_{(12)}^{(2)})^{n_p(g)} \overrightarrow{X}^{sg\overline{s}}
    (Z_{(01)}^{(1)})^{n_q(h)} \overrightarrow{X}^{th\overline{t}}\overrightarrow{X}^{ugh\overline{u}}\dots \\
    &= (-1)^{n_p(g)n_q(h)}\times \dots(Z_{(12)}^{(2)})^{n_p(g)}(Z_{(01)}^{(1)})^{n_q(h)} \overrightarrow{X}^{sg\overline{s}}\overrightarrow{X}^{th\overline{t}}\overrightarrow{X}^{ugh\overline{u}}\dots \\
    &= \exp(i\omega_{pq}(g,h))\times \dots(Z_{(12)}^{(2)})^{n_p(g)}(Z_{(01)}^{(1)})^{n_q(h)} \overrightarrow{X}^{sg\overline{s}}\overrightarrow{X}^{th\overline{t}}\overrightarrow{X}^{ugh\overline{u}}\dots
\end{split}
\end{align}
so by moving the Pauli $Z$ operator to the front of the expression, we get the phase factor $\exp(i\omega_{pq}(g,h))$ from each tri-junction of the magnetic fluxes. Such a phase factor arises from each junction, so the action on the ribbon network is given by
\begin{align}
    \mathcal{R}_{\Z_2} K \mathcal{R}^\dagger_{\Z_2} = \left(\prod_{\text{junctions}} \exp( i\omega_{pq}(g,h)) \right) K~,
\end{align}
where the product is over the junctions of magnetic fluxes. 
We note that since $[\omega_{pq}] = [\omega_{\text{sf}}]$ in cohomology $H^2(B\gsf,U(1))$, the above action on the closed ribbon network can be expressed as
\begin{align}
    \mathcal{R}_{\Z_2} K \mathcal{R}^\dagger_{\Z_2} = \left(\prod_{\text{junctions}} \exp( i\omega_{\text{sf}}(g,h)) \right) K~,
\end{align}
This is consistent with the $U$ symbol of the symmetry fractionalization data provided in \eqref{eq:Usymbolnew}.

\subsection{Emergent symmetry and faithful action on genus $g \geq 2$ surfaces}

Each ground state of the quantum double model on a genus $g$ surface can be written as a superposition over gauge-equivalent flat $G$ gauge field configurations on the 2d space. Different ground states correspond to gauge-inequivalent gauge field configurations. The operator $\mathcal{R}_{\Z_2}$ acts on each configuration of the gauge field by a phase $\exp(i\int\omega_{\text{sf}})$. Since the integral gives a gauge invariant partition function of the SPT, this implies that $\mathcal{R}_{\Z_2}$ acts on each state by a phase factor. Since the integral $\int \omega_{\text{sf}}$ vanishes on the $\gsf$ gauge field on a torus, the symmetry operator $\mathcal{R}_{\Z_2}$ acts as the identity operator on the ground state subspace on a torus. 

\begin{figure}[t]
    \centering
    \includegraphics[width=0.5\textwidth]{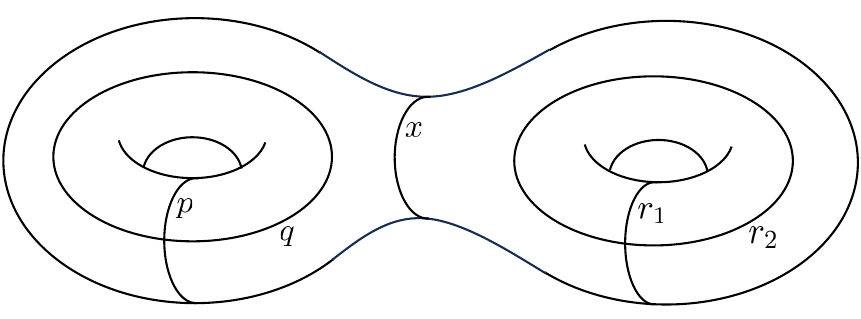}
    \caption{The holonomy of $\Z_2$ gauge fields on a genus 2 surface on which $\int \omega_{\text{sf}}$ is nontrivial.}
    \label{fig:genus2}
\end{figure}

Meanwhile, $\mathcal{R}_{\mathbb{Z}_2}$ defines a non-trivial operator on the ground state subspace of a genus two surface. For instance, let us consider the genus two surface where  each pair of cycles carries the $\gsf$ holonomy $(p,q)$, $(r_1,r_2)$. Since $pqp^{-1}q^{-1} = r_1r_2 r_1^{-1}r_2^{-1}=x$, this defines a nontrivial $\gsf$ gauge field. See Fig.~\ref{fig:genus2}.
The operator $\mathcal{R}_{\Z_2}$ then evaluates to a $(-1)$ phase on the state with this $\gsf$ gauge field configuration. In general, the symmetry action on the Hilbert space is a diagonal matrix with elements $1$ or $-1$ which are partition functions of the (1+1)D $\omega_{\text{sf}}$ SPT with the given $\gsf$ background. 

\section{Distinct gapped boundaries with identical condensed particles}
\label{sec:boundary}

The gapped boundary condition of (2+1)D finite $G$ gauge theory is labeled by a pair $(K,\omega)$; the choice of a subgroup $K\subset G$ and the 2nd group cohomology $\omega\in H^2(K,U(1))$~\cite{Beigi2011boundary}. Here we consider two distinct gapped boundaries of the untwisted $\gsf$ gauge theory labeled by $(\gsf,0), (\gsf,\omega_{\text{sf}})$. These two gapped boundaries correspond to condensation of the magnetic particles, with or without the $\Z_2$ symmetry operator $\omega_{\text{sf}}$ support at the boundary.
Since the defect $\omega_{\text{sf}}$ does not permute anyons, the two gapped boundaries have the same set of condensed particles. In other words, these two gapped boundaries have the same Lagrangian algebra object $\mathcal{L}$ given by 
\begin{align}
    \mathcal{L} = \bigoplus_{[g]\in Cl(\gsf)} ([g],1)
\end{align}

However, the set of condensed particles does not uniquely determine the gapped boundary condition. The gapped boundary is characterized by the Lagrangian algebra, which takes the Lagrangian algebra object $\mathcal{L}$ and the multiplication morphism $\mu: \mathcal{L}\otimes \mathcal{L}\to\mathcal{L}$ \cite{davydov2011wittgroupnondegeneratebraided,Fuchs2013lagrangian}. Due to the nontrivial action of the defect $\omega_{\text{sf}}$ on the fusion vertices of the magnetic defects, the above two gapped boundaries have the distinct multiplication morphism $\mu$ that leads to the distinct algebraic structure.\footnote{We thank Sahand Seifnashri for helpful discussions on this point.}

Let us comment about the characterization of the Lagrangian algebra in terms of modularity. It is known that the Lagrangian algebra object $\mathcal{L}$ satisfies the modularity condition; suppose we have $\mathcal{L}=\bigoplus_a Z_{0a} a$ with simple anyons $a$ and non-negative coefficients $\{Z_{0a}\}$, then the coefficients of the Lagrangian algebra object satisfy $SZ=Z, TZ=Z$ with modular $S,T$ matrices~\cite{Lan2015gapped}. These equations are derived by putting the gapped boundary on the torus, and studying the modular invariance of the corresponding boundary state~\cite{Kaidi2022boundary}. However, since the above conditions do not rely on the multiplication morphism $\mu$, this does not give a faithful characterization of the Lagrangian algebra; it does not distinguish the gapped boundaries $(\gsf,0)$ and $(\gsf,\omega_{\text{sf}})$. Even before that, it is known that the above modularity condition is not a sufficient condition for the Lagrangian algebra object $\mathcal{L}$ \cite{Kawahigashi_2015}.
One can extend the above modularity conditions to the higher-genus case by considering the gapped boundary at generic Riemann surface, and studying the invariance of the corresponding boundary state under mapping class groups~\cite{Kaidi2022boundary}.~\footnote{Here, the boundary state is constructed by considering the TQFT on $\Sigma_g\times I$ where $\Sigma_g$ is the genus $g$ surface and $I$ is the interval, and then one end of the interval is filled with the gapped boundary. This defines a TQFT state on $\Sigma_g$ on the other end, which is the boundary state. Due to the topological nature of the gapped boundary, the boundary state has to be invariant under the mapping class group actions of $\Sigma_g$.} However, such family of conditions within the genus $< g$ still does not fully characterize the Lagrangian algebra. For instance, there are (1+1)D SPT phases with a certain finite group $G^{(g)}_{\text{sf}}$ which cannot be detected on the genus $<g$ surface, but can be detected on genus $g$ (see Appendix \ref{app:highergenus}). This leads to a pair of gapped boundaries of the $G_{\text{sf}}^{(g)}$ gauge theory $(G_{\text{sf}}^{(g)},0),(G_{\text{sf}}^{(g)},\omega)$ that cannot be distinguished from the boundary state with genus $<g$. 

\subsection{Application: Oblique SSB phase of $\mathrm{Rep}(G_{\mathrm{sf}})$ symmetry in (1+1)D}

The (2+1)D topological order and its gapped boundary are used to describe the gapped phases in (1+1)D through the symmetry TQFT \cite{Ji:2019jhk,Freed:2022qnc,Kaidi:2022cpf, bhardwaj2023categoricallandauparadigmgapped, bhardwaj2024gappedphasesnoninvertiblesymmetries}. Namely, the (1+1)D gapped phase is regarded as a (2+1)D topological order on a thin interval, and then each end of the interval is realized by a gapped boundary. The topological operators on either side of the gapped boundary generates the global symmetry of the system.

Intriguingly, the almost identical pair of gapped boundaries described above leads to the two distinct spontaneous broken phases of $\mathrm{Rep}(G_{\text{sf}})$ symmetry, where the symmetry is completely broken. One SSB phase is given by the interval of the (2+1)D $G_{\text{sf}}$ gauge theory sandwiched by the same gapped boundaries $(G_{\text{sf}},0),(G_{\text{sf}},0)$. The other SSB phase is given by the interval sandwiched by two distinct gapped boundaries $(G_{\text{sf}},0),(G_{\text{sf}},\omega_{pq})$. 

These two phases have the identical set of 0-form $\mathrm{Rep}(G_{\text{sf}})$ symmetry operators and the same local (topological) operators. The 0-form symmetry operators are described by the electric Wilson line operators of the $G_{\text{sf}}$ gauge theory stretching parallel to the interval. Meanwhile, the local operators are described by the magnetic defects stretching between the ends of the interval. The $\text{Rep}(\gsf)$ symmetry operator acts on the local operators by the mutual braiding between the electric Wilson lines and the magnetic flux.
Each of the $\text{Rep}(G_{\text{sf}})$ symmetry defects nontrivially acts on at least one of the local operators, and the $\text{Rep}(G_{\text{sf}})$ is maximally broken in both phases.

Nevertheless, these two phases with fully broken $\text{Rep}(G_{\text{sf}})$ are distinct phases. In particular, if one SSB phase is continuously connected to the other by a deformation preserving $\text{Rep}(G_{\text{sf}})$ symmetry, it has to go through a phase transition. This can be most easily seen by gauging the $\text{Rep}(G_{\text{sf}})$ symmetry~\cite{Bhardwaj2018gauging}, then the path connecting the SSB phases becomes the path from a trivial to nontrivial SPT phase preserving the dual $G_{\text{sf}}$ symmetry. This has to experience a phase transition.\footnote{We thank Nat Tantivasadakarn for helpful discussions on this point.} See \cite{kobayashi2025projective} for explicit lattice models of distinct SSB phases of $\text{Rep}(G_{\text{sf}})$ symmetry.

\section{Higher dimensions: $Q_8$ gauge theory in (3+1)D}
\label{sec:3+1d}
We have seen that there are soft global symmetries in (2+1)D topological orders that do not permute anyons, but nontrivially acts on the junction of the anyons. In this section, we discuss an analogue of soft symmetries of topological order in higher dimensions. 

Typically, it becomes easier to find such non-permuting symmetries in higher dimensions. For instance, in a standard (3+1)D $\Z_2$ gauge theory with a bosonic electric particle, there is a $\Z_2$ 0-form symmetry that corresponds to the (2+1)D $\Z_2$ Levin-Gu SPT phase. This SPT has slant product which is trivial in cohomology, hence leads to the trivial permutation action on the magnetic fluxes.
However, this $\Z_2$ symmetry still has the nontrivial algebraic mixture with the magnetic fluxes and Wilson lines, which together forms a three-group symmetry. For instance, the junction of  $\Z_2$ 0-form symmetry operators acts on the magnetic surface operator by attaching an electric Wilson line operator~\cite{Barkeshli2024highergroup, Hsin2020spinliquids}.

Therefore, a nontrivial question is whether one can find a faithful 0-form symmetry action in (3+1)D topological order, which does not permute any topological operators, and forms trivial higher-group symmetry with other topological operators. 

One can find such 0-form global symmetry in (3+1)D gauge theory of a quaternion group $Q_8$. Let us consider the global symmetry that corresponds to the (2+1)D $Q_8$ SPT phase, classified by $H^3(BQ_8,U(1))= \Z_8$. This generates the faithful $\Z_8$ symmetry of the $Q_8$ gauge theory.
We will see that the $\Z_2$ subgroup of this $\Z_8$ symmetry has the trivial higher group structure with other topological operators, meaning that the $\Z_2\subset \Z_8$ becomes a direct product with other symmetries of $Q_8$ gauge theory.

\begin{figure}[t]
    \centering
    \includegraphics[width=0.7\textwidth]{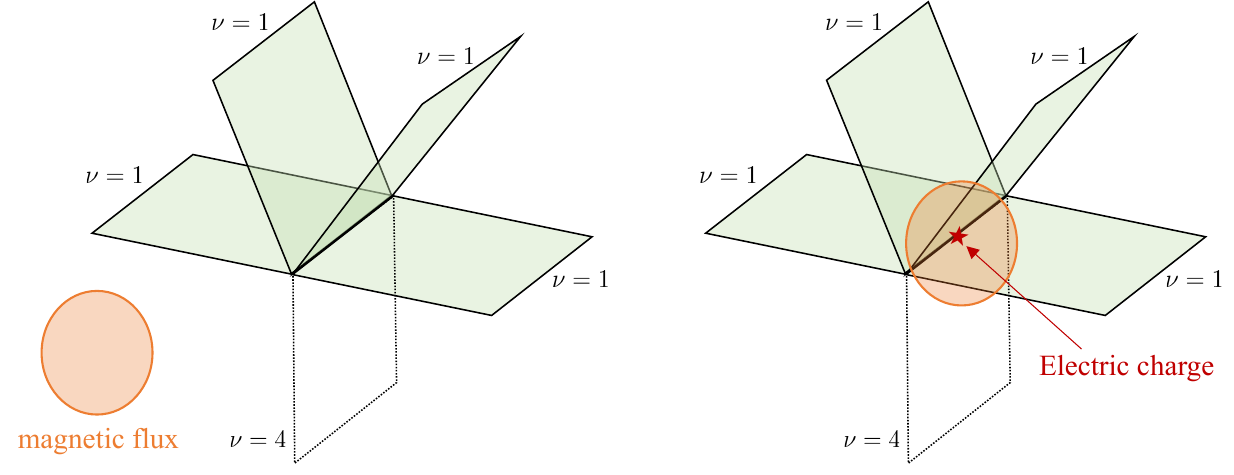}
    \caption{The junction of four $\nu=1$ defects into a $\nu=4$ defect. When the non-invertible magnetic surface operator $[\mathbf{i}]=[\mathbf{j}]=[\mathbf{k}]$ intersects with this junction, their intersection emits an electric charge carried by the 2d Wilson line of $Q_8$ gauge theory.}
    \label{fig:n1junctions}
\end{figure}

Let us take a $Q_8$ SPT phase that generates the $\Z_8$ classification, labeled by $\nu=1\in\Z_8$. We write the generators of $Q_8$ (quaternions) as $\mathbf{i},\mathbf{j},\mathbf{k}$, satisfying $\mathbf{i}^2=\mathbf{j}^2=\mathbf{k}^2=-1$. Then, if we restrict the gauge group to the $\Z_4$ subgroup $\{1,\mathbf{i}, -1,-\mathbf{i}\}$, this $\nu=1$ SPT phase reduces to the $\Z_4$ SPT that generates the $H^3(B\Z_4,U(1))=\Z_4$ classification~\cite{Yu2022genusone}. Therefore, writing the topological response of the $\nu=1$ phase as $\omega_{\nu=1}\in Z^3(BQ_8,U(1))$, its restriction to the $\Z_4$ gauge field becomes $\omega_{\nu=1} = \frac{2\pi i}{4} a \frac{d\hat{a}}{4}$, where $a$ is the $\Z_4$ gauge field for $\Z_4=Z(\mathbf{i})=\{1,\mathbf{i}, -1,-\mathbf{i}\}$, and $\hat{a}$ is the lift to $\Z_{16}$.
Its slant product with respect to $\mathbf{i}\in Q_8$ is evaluated as 
\begin{align}
    i_{\mathbf{i}}(\omega_{\nu=1}) = \frac{2\pi i}{4}\frac{d\hat{a}}{4}
\end{align}
Since this slant product is trivial in cohomology $H^2(B\Z_4,U(1))$, this $\nu=1$ phase does not induce permutation of the magnetic flux $[\mathbf{i}$]. However, the above slant product leads to a subtle algebraic mixture among SPT defects, the magnetic fluxes and Wilson lines~\cite{Barkeshli2024highergroup}. Suppose we have the junction of four SPT defects labeled by $\nu=1$ into $\nu=4$. If the magnetic surface operator labeled by the conjugacy class $[\mathbf{i}]$ intersects with this junction, the intersection has to emit a $\Z_4$ electric Wilson line $\exp(\frac{2\pi i}{4} \int a)$ carrying the charge of $\Z_4=Z(\mathbf{i})$. Since only the 2d irreducible representation of $Q_8$ can carry the $\Z_4$ charge under $Z(\mathbf{i})$, one can say that this junction emits the 2d Wilson line of $Q_8$.
This is a non-invertible categorical symmetry that involves the junction of 0-form SPT defects, magnetic fluxes and the 2d non-invertible Wilson line. See Fig.~\ref{fig:n1junctions} for illustrations.
The whole discussion about the $\nu=1$ defect above is symmetric under permuting group elements among $\mathbf{i},\mathbf{j},\mathbf{k}$.

Now let us restrict the 0-form symmetry to the $\Z_2\subset \Z_8$ global symmetry generated by $\nu=4$ phase.\footnote{Though this $\Z_2\subset \Z_8$ symmetry is faithful, it acts on the Hilbert space in a quite subtle manner. In particular, its action on the Hilbert space at any mapping torus of $T^2$ becomes trivial. This $\Z_2$ symmetry acts faithfully on the Hilbert space at e.g., a spherical manifold $S^3/Q_8$ dubbed a prism manifold~\cite{Yu2022genusone}.} The slant product for any group element becomes trivial,
\begin{align}
    i_{\mathbf{i}}(\omega_{\nu=4}) = i_{\mathbf{j}}(\omega_{\nu=4}) = i_{\mathbf{k}}(\omega_{\nu=4}) = 0.
\end{align}

\begin{figure}[t]
    \centering
    \includegraphics[width=0.8\textwidth]{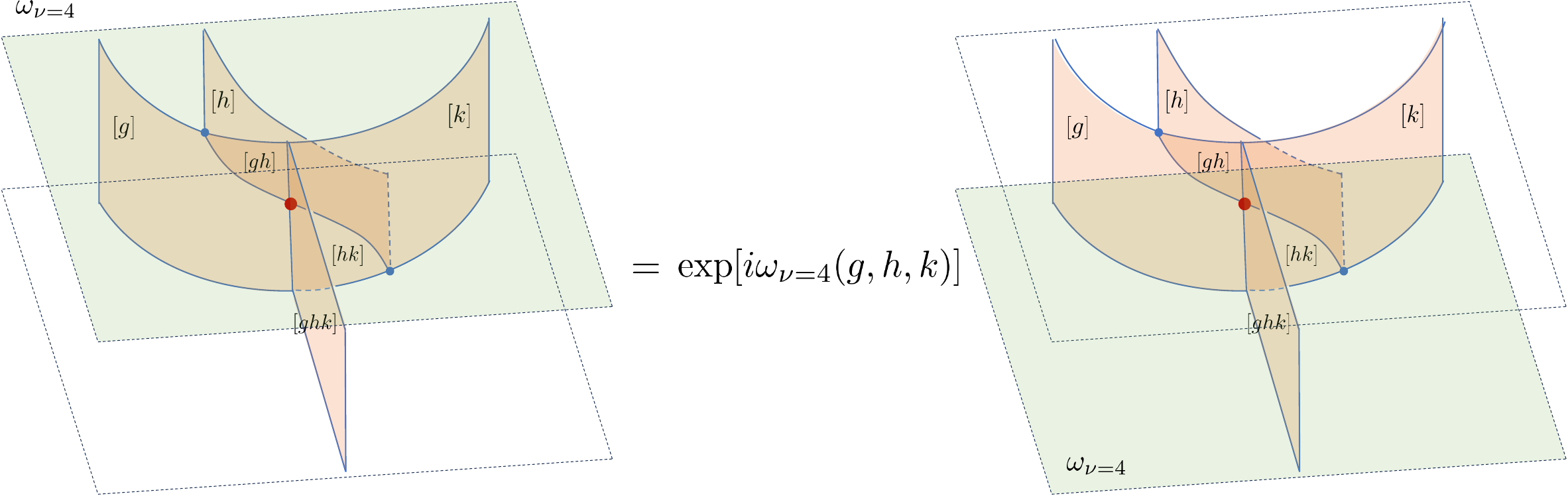}
    \caption{The $\nu=4$ defect acts on the point junction of three magnetic fluxes by the phase factor.}
    \label{fig:softQ8}
\end{figure}

This implies that the $\nu=4$ defect does not permute the topological operators, and becomes direct product with the symmetries of magnetic fluxes and Wilson lines.
However, this $\nu=4$ symmetry still acts by phase factors on the point junction of the three magnetic surface operators, see Fig.~\ref{fig:softQ8}. 
This is an analogue of the soft symmetry of (2+1)D topological orders.

More generally, when $G$ is a discrete subgroup of $SU(2)$, one can always find a topological response $\omega\in Z^3(BG,U(1))$ which is nontrivial in cohomology $[\omega]\neq 0$, but its slant product with any group element $g\in G$ becomes completely trivial $i_g\omega=0$ at the cochain level~\cite{Yu2022genusone}.
This implies that this class of finite groups has an analogue of soft symmetry, whose only nontrivial feature is an action on the point junctions of magnetic surfaces by a phase factor.

\section{Discussion}
\label{sec:discussions}

In this paper we have demonstrated the existence of soft symmetries in (2+1)D and (3+1)D topological orders. In (2+1)D, these correspond to symmetries that are neither fractionalized nor permute the anyons, but which nevertheless have a non-trivial action on the topological ground state subspace on higher genus surfaces. In (3+1)D, they correspond to symmetries that do not permute any topological defects and define a completely trivial 3-group structure with the other topological defects, but nevertheless act non-trivially in the underlying topological quantum field theory. 

For the (2+1)D theory, we studied an explicit lattice model realization, where the soft symmetry is realized by pumping a gauged (1+1)D SPT through the system via a finite depth local unitary circuit. Such a circuit does not commute with the full Hamiltonian but keeps the ground state subspace invariant, and thus can be viewed as an emergent symmetry of the theory. 

So far we have demonstrated how the quantum double lattice model possesses an operator that acts as a non-trivial unitary operator on the ground state subspace. As such, it defines an emergent non-onsite symmetry. A natural question is whether we can construct a lattice model in which it corresponds to an exact, on-site symmetry. 

One possible method for achieving this on-site soft symmetry is to use the symmetry-enriched version of the Crane-Yetter-Walker-Wang construction presented in \cite{Bulmash2020absolute}. 

Ref. \cite{Bulmash2020absolute} provides the (3+1)D path integral state sum model and corresponding Hamiltonian of the (3+1)D bosonic invertible phase with $G$ symmetry, with input being a UMTC and its $G$ symmetry fractionalization data $\{U,\eta\}$. This (3+1)D theory has a surface SET phase on its (2+1)D boundary, described by the input UMTC enriched with $G$ symmetry. 
It would be interesting to see if the construction of the path integral is also valid for the symmetry fractionalization data $\{U,\eta\}$ for the soft symmetry discussed in this paper. 

Once one can construct the (3+1)D SPT by the Walker-Wang model with the boundary realizing SET with the soft symmetry, we note that the bulk (3+1)D phase is necessarily topologically trivial, because there are no non-trivial $\mathbb{Z}_2$ SPTs in (3+1)D \cite{chen2011,kapustin2014SPTbeyond}. Since the bulk (3+1)D phase is topologically trivial, there should also be a symmetry-preserving topologically trivial boundary condition. Considering the system on a thin slab with the top boundary corresponding to our SET of interest and the bottom slab corresponding to the topologically trivial boundary condition should give us a (2+1)D lattice model where the $\mathbb{Z}_2$ soft symmetry is an exact on-site symmetry. It would be interesting to explicitly construct such (2+1)D lattice models with on-site soft symmetries.

Also, the soft symmetries that we studied in this paper were in terms of gauged SPT defects. A natural question is whether there is a systematic method to find all possible soft symmetries. For example, as discussed in Sec. \ref{sec:softexample_gen}, there are also soft symmetries of $G$ gauge theory corresponding to Out$_{\text{sf}}(G)$. It would be interesting to develop a physical construction for such soft symmetries that allows them to be implemented as a finite depth circuit. 

Finally, it remains unknown if the generic soft symmetry has the $H^3$ obstruction characterized by $[\mathcal{O}_3] = \mathcal{H}^3(\text{Aut}_{br}(\mathcal{C}),\mathcal{A})$. While the example of the soft symmetry provided in the main text does not have the $H^3$ obstruction, it remains unclear if the generic soft symmetries including the outer automorphism $\text{Out}_{\text{sf}}(G)$ carry the trivial $H^3$ obstruction or not.\footnote{We note that \cite{Benini2018} argued that the $H^3$ obstruction must be trivial when the symmetry does not permute anyons. Their argument assumes that the fusion multiplicities satisfy $N^{ab}_c\in\{0,1\}$ for all fusion vertices, and that the $F$-symbol $(F^{abc}_{d})_{ef}$ is always nonzero whenever the four relevant fusion spaces $V^{ab}_e$, $V^{ec}_d$, $V^{af}_d$, and $V^{bc}_f$ are all nonempty. We note, however, that there are modular tensor categories in which certain $F$-symbols vanish even though this is not forbidden by the fusion rules; for example, $F^{111}_1$ vanishes in $SU(2)_4$. Therefore, it is still unknown in general if $H^3$ obstruction is vanishing for soft symmetries. We thank Parsa Bonderson for bringing this point to our attention.
 } See Refs.~\cite{Barkeshli2019, Fidkowski2017H3} for examples of the outer automorphism in finite gauge theories with $H^3$ obstructions.
This is an interesting question since it is often believed that the symmetry that does not permute anyons leads to trivial $H^3$ obstruction \cite{Barkeshli2019, Benini2018}.

\section{Acknowledgment}

We thank Parsa Bonderson, Yu-An Chen, Meng Cheng, Yichul Choi, Po-Shen Hsin, Yoshiko Ogata, Abhinav Prem, Nathan Seiberg, Sahand Seifnashri, Nikita Sopenko, Nat Tantivasadakarn and Zhenghan Wang for useful discussions. We thank Meng Cheng, Nat Tantivasadakarn and Zhenghan Wang for comments on the draft.
R.K. is supported by the U.S. Department of Energy (Grant No.~DE-SC0009988) and the Sivian Fund. M.B. is supported by NSF DMR-2345644. 

\appendix

\section{Symmetric representative of second cohomology $H^2(B\gsf,U(1))$}
\label{app:cocycle}
In the main text, we provided the representative of $H^2(B\gsf,U(1))$ given by the following 2-cocycle
\begin{align}
    \omega_{pq}(g,h) = \pi (n_p(g)n_q(h))~.
\end{align}
Here we explicitly construct a function $\chi: \gsf\to\Z_2$ such that
\begin{align}
    \omega_{\text{sf}}(g,h) := \omega_{pq}(g,h)+\pi(\chi(g)+\chi(h)-\chi(gh))
\end{align}
satisfies the following two properties:
\begin{itemize}
    \item $\omega_{\text{sf}}(g,h)$ is symmetric, i.e., $\omega_{\text{sf}}(g,h)=\omega_{\text{sf}}(h,g)$ for any $g,h\in\gsf$.
    \item $\omega_{\text{sf}}(g,h)$ is invariant under overall conjugation by group elements, i.e., $\omega_{\text{sf}}(g,h) = \omega_{\text{sf}}(kgk^{-1},khk^{-1})$ for any $g,h,k\in\gsf$.
\end{itemize}
Note that $\omega_{\text{sf}}$ gives another representative of $H^2(B\gsf,U(1))$ with $[\omega_{\text{sf}}]=[\omega_{pq}]$.
Let us explicitly construct such a function $\chi$.
For convenience, a group element $g\in\gsf$ is sometimes expressed as
$g=g_A g_B$, where $g_A=p^\alpha q^\beta r_1^{\gamma_1} r_2^{\gamma_2}$ and $g_B = x^{\mu} y_1^{\nu_1}y_2^{\nu_2}$. We define a subset $\Gamma_A\subset \gsf$ that consists of group elements $\Gamma_A=\{p^\alpha q^\beta r_1^{\gamma_1} r_2^{\gamma_2}\}$ with generic choices of $\alpha,\beta,\gamma_1,\gamma_2$. Let us similarly define the subgroup $\Gamma_B \subset \gsf$ where $\Gamma_B = \{x^{\mu} y_1^{\nu_1}y_2^{\nu_2}\}$.

We first fix $\chi(g)$ for the group elements $g\in\gsf$ expressed as $g=kg_Ak^{-1}$ for some $g_A\in\Gamma_A, k\in\gsf$. We set
\begin{align}
    \chi(kg_A k^{-1}) = n_p(g_A) n_q(k) + n_p(k) n_q(g_A) \quad \mod 2.
\end{align}

Not all group elements $g\in\gsf$ admits an expression $g=kg_Ak^{-1}$ (for instance, $g=qy_1$ does not).
For a generic group element $g\in\gsf$, we set $\chi(g)$ in the following fashion. 
For $g_A\in\Gamma_A$, we define a group $\Gamma(g_A)\subset \Gamma_B$, which is the group generated by the elements $\gamma_k:=kg_Ak^{-1}(g_A)^{-1}$ for all possible choices of $k\in\gsf$. Then, let us fix a choice of another subgroup $\tilde\Gamma(g_A)\subset \Gamma_B$ satisfying $\Gamma(g_A)\times\tilde\Gamma(g_A)=\Gamma_B, \Gamma(g_A)\cap\tilde{\Gamma}(g_A) = \{\text{id}\}$. 

Then, $\chi(g)$ with generic group element $g\in\gsf$ is defined as follows. We express $g= kg_A\tilde\gamma k^{-1}$ with $g_A\in\gsf, \tilde\gamma\in\tilde\Gamma(g_A), k\in\gsf$. By definition, for a given $g\in\gsf$ the choice of $g_A,\tilde\gamma$ are uniquely fixed.
We then set
\begin{align}
    \chi(kg_A\tilde\gamma k^{-1}) = n_p(g_A) n_q(k) + n_p(k) n_q(g_A) \quad \mod 2.
    \label{eq:chidef}
\end{align}
For a given $g\in\gsf$, one can explicitly check that the above $\chi(g)$ does not depend on the choice of $k\in\gsf$. 

Let us confirm that the properties of $\omega_{\text{sf}}(g,h)$ are satisfied with the above choice of $\chi$.
Let us first check the symmetric property. We have
\begin{align}
    e^{i\omega_{\text{sf}}(g,h)-i\omega_{\text{sf}}(h,g)} = (-1)^{n_p(g)n_q(h) - n_q(g)n_p(h) + \chi(gh) - \chi(hg)}~.
\end{align}
Let us write $gh = k(gh)_A \tilde \gamma k^{-1}$ with $(gh)_A\in\Gamma_A, \tilde \gamma\in \tilde\Gamma((gh)_A)$, $k\in\gsf$. Then we get
\begin{align}
    \chi(gh) = n_p((gh)_A)n_q(k) + n_p(k)n_q((gh)_A), \quad \chi(hg) = n_p((gh)_A)n_q(hk) + n_p(hk)n_q((gh)_A)~.
\end{align}
These can be rewritten as
\begin{align}
\begin{split}
    \chi(gh) &= (n_p(g)+n_p(h))n_q(k) + n_p(k)(n_q(g)+n_q(h)), \\
    \chi(hg) &= (n_p(g)+n_p(h))(n_q(h)+n_q(k)) + (n_p(h)+n_p(k))(n_q(g)+n_q(h))
    \end{split}
\end{align}
Hence we have $\chi(gh)-\chi(hg) =n_p(g)n_q(h) - n_q(g)n_p(h)$ mod 2. By plugging this to the expression of $e^{i\omega_{\text{sf}}(g,h)-i\omega_{\text{sf}}(h,g)}$, one can see that $\omega_{\text{sf}}(g,h)$ is symmetric.

Let us then verify the property $\omega_{\text{sf}}(g,h) = \omega_{\text{sf}}(kgk^{-1},khk^{-1})$ for any $g,h,k\in\gsf$.
To see this, one can first immediately check that $\omega_{pq}$ satisfies $\omega_{pq}(g,h) = \omega_{pq}(kgk^{-1},khk^{-1})$. So one just needs to check
\begin{align}
    d\chi(g,h) = d\chi(kgk^{-1},khk^{-1})~.
\end{align}
From the definition of $\chi$ in \eqref{eq:chidef}, for generic $g,k\in\gsf$ we have
\begin{align}
    \chi(g)-\chi(kgk^{-1}) = n_p(g)n_q(k) + n_p(k)n_q(g)\quad\mod 2.
\end{align}
Therefore we have
\begin{align}
    d\chi(g,h) - d\chi(kgk^{-1},khk^{-1}) = dn_p(g,h)n_q(k) + n_p(k) dn_q(g,h) = 0 \quad \mod 2.
\end{align}
This completes the proof of $\omega_{\text{sf}}(g,h) = \omega_{\text{sf}}(kgk^{-1},khk^{-1})$.

\section{(1+1)D SPT phases with genus $\ge 3$ detection}
\label{app:highergenus}

In this appendix, we construct the (1+1)D SPT phases which can be detected on a genus $g$ surface, though cannot be detected on genus $\le g-1$.

We consider the group $G^{(g)}_{\text{sf}}$ of order $2^{g^2+2g-1}$ with $g\ge 2$ which is the central extension $\Z_2^{g^2-1}\to G^{(g)}_{\text{sf}}\to \Z_2^{2g}$. The generators of $\Z_2^{2g}$ are labeled by $p_j,q_j$ with $1\le j\le g$, and generators of $\Z_2^{g^2-1}$ are labeled by $x_j$ with $1\le j\le g-1$, and $y_{jk},z_{jk}$ with $1\le j<k \le g$. The central extension is characterized by
\begin{align}
    \begin{split}
        [p_1,q_1]&=x_1, \quad [p_g,q_g]=x_{g-1}, \quad \\
        [p_j,q_j] &=x_{j-1}x_j; \ 2\le j \le g-1 \\
        [p_j,p_k] &= y_{jk};\  j<k \\
        [p_j,q_k] &= z_{jk};\  j<k
    \end{split}
\end{align}
where $[p,q]=pqp^{-1}q^{-1}$ is a group commutator. Note that the above group $G^{(g)}_{\text{sf}}$ with $g=2$ corresponds to the group $\gsf$ discussed in the main text.
Let us consider the SPT action $\omega^{(g)}\in Z^2(BG^{(g)}_{\text{sf}},U(1))$ characterized by
\begin{align}
    \omega^{(g)}(h,k) = \pi (n_{p_1}(h) \cup n_{q_1}(k)),
\end{align}
where $n_{p_1}(h)\in Z^1(BG^{(g)}_{\text{sf}},U(1))$ is the mod 2 number of $p_1$ that appears in the expression of $h\in G^{(g)}_{\text{sf}}$ in terms of the product of group generators. Similar for $n_{q_1}(h)$. This SPT phase is detected on a genus $g$ surface with holonomy $(p_j,q_j)$, $1\le j\le g$ on each pair of cycles, while cannot be detected on closed oriented surfaces of genus $<g$. 

The above SPT phases lead to the ``softer'' 0-form $\Z_2$ symmetry of (2+1)D $G^{(g)}_{\text{sf}}$ gauge theory with nontrivial action on genus $g$, generated by the gauged SPT defect.

\bibliographystyle{utphys}
\bibliography{bibliography,bib}

\end{document}